\documentclass{aa}  

\usepackage{graphicx}
\usepackage{txfonts}

\usepackage{xcolor}

\newcommand{\ebv}{$E(B-V)$}

\newcommand{\hb}{H$\beta$}
\newcommand{\ha}{H$\alpha$}

\newcommand{\oi}{[O\,\textsc{i}]}

\begin{document} 

   \title{Diffuse interstellar bands $\lambda$5780 and $\lambda$5797 \\in the Antennae Galaxy as seen by MUSE}

\author{A. Monreal-Ibero
    \inst{1,2}
    \and
    P.~M.\ Weilbacher\inst{3}
    \and 
    M. Wendt\inst{3,4}
}
    \institute{
    Instituto de Astrof\'{\i}sica de Canarias (IAC), E-38205 La Laguna, Tenerife, Spain
    \and
    Universidad de La Laguna, Dpto.\ Astrof\'{\i}sica, E-38206 La Laguna, Tenerife, Spain\\
    \email{amonreal@iac.es}
    \and 
    Leibniz-Institut f\"ur Astrophysik Potsdam (AIP), An der Sternwarte 16, D-14482 Potsdam, Germany
    \and 
    Institut f\"ur Physik und Astronomie, Universit\"at Potsdam,Karl-Liebknecht-Str. 24/25, 14476 Golm, Germany
}

   \date{Received October 26, 2017; accepted XXXX YY, 201Z}

 
  \abstract
   {Diffuse interstellar bands (DIBs) are faint spectral absorption features of unknown origin. Research on DIBs beyond the Local Group is very limited and will surely blossom in the era of the Extremely Large Telescopes. However, we can already start paving the way. One possibility that needs to be explored is the use of high-sensitivity integral field spectrographs.}
   {Our goals are twofold. First, we aim to derive reliable mapping of at least one  DIB in a galaxy outside the Local Group. Second, we want to explore the relation between DIBs and other properties of the interstellar medium (ISM)  in the galaxy.
}
   {We use Multi Unit Spectroscopic Explorer (MUSE) data for the Antennae Galaxy, the closest major galaxy merger. High signal-to-noise spectra were created by co-adding the signal of many spatial elements with the Voronoi binning technique. The emission of the underlying stellar population was modelled and substracted with the \texttt{STARLIGHT} spectral synthesis code. Flux and equivalent width of the features of interest were measured by means of fitting to Gaussian functions.}
   {
To our knowledge, we have derived the first maps for the DIBs at $\lambda$5780 and $\lambda$5797 in galaxies outside the Local Group.
The strongest of the two DIBs (at $\lambda$5780) was detected in an area of $\sim0.6\square^\prime$, corresponding to a linear scale of $\sim25$~kpc$^2$. 
This region was sampled using $>$200 out of $\sim$1200 independent lines of sight.
The DIB $\lambda$5797 was detected in $>$100 independent lines of sight.
Both DIBs are associated with a region of high emission in the \ion{H}{i} 21~cm line, implying a connection between atomic gas and DIBs, as the correlations in the Milky Way also suggest.
Conversely, there is mild spatial association between the two DIBs and the molecular gas, in agreement with results for
our Galaxy that indicate a lack of correlation between DIBs and molecular gas.
The overall structures for the DIB strength distribution and extinction are comparable. Within the system, the $\lambda$5780 DIB clearly correlates with the extinction, and both DIBs follow the relationship between equivalent width and reddening when data for several galaxies are considered. This relationship is tighter when comparing only with galaxies with metallicities close to solar.
Unidentified infrared emission bands (UIBs, likely caused by polycyclic aromatic hydrocarbons (PAHs)) and the $\lambda$5780 and $\lambda$5797 DIBs show similar but not identical spatial distributions. We attribute the differences to extinction effects without necessarily implying a radically different nature of the respective carriers.
}
{The results illustrate the enormous potential of integral field spectrographs for extragalactic DIB research.}

   \maketitle
%

\section{Introduction}

Stellar spectra display a plethora of weak absorption features called diffuse interstellar bands \citep[DIBs, see][for reviews]{Herbig95,Sarre06}. These were first identified by \citet{Heger22} and their interstellar origin was established a few years later \citep{Merrill36}. Today, we know more than 400 optical DIBs \citep[e.g.][]{Jenniskens94,Hobbs09}. Additionally, $\sim$30 DIBs have been detected in the near-infrared \citep[e.g.][]{Joblin90,Geballe11,Cox14,Hamano16,Elyajouri17} while there is no firm detection in the near-UV yet \citep{Bhatt15}.

Identification of the carriers (i.e.\ molecules causing the absorptions) has proven to be an arduous task.
With the exception of $C^+_{60}$, recently confirmed as the carrier of the DIBs at $\lambda$9577 and $\lambda$9632 \citep[][but also see \citealt{Galazutdinov17} for objections to this identification]{Campbell15,Walker15}, those for the vast majority of DIBs remain unidentified. Hydrocarbon chains \citep{Maier04,Motylewski00,Krewlowski11}, polycyclic aromatic hydrocarbons \citep[PAHs, ][]{Leger85,vanderZwet85,Salama96}, and fullerenes \citep{Kroto88,Sassara01,IglesiasGroth07} are among the possible candidates. In general, carbon seems to be involved and in that sense, DIBs have been proposed as the largest reservoir of organic matter in the Universe \citep{Snow14}.

The number of viable candidates can be reduced by studying the DIBs' intrinsic characteristics (e.g.\ band profiles) and the relationships between different DIBs and the interstellar medium (ISM), in general. Thus, it was found that most of the strong DIBs are correlated \citep[e.g.][]{Friedman11,Kos13,Puspitarini13}, though this is not always the case for weaker DIBs, which can even show a negative correlation \citep{Baron15b}. However, with the possible exception of the $\lambda$6196.0-$\lambda$6613.6 pair, none of the derived correlations can be considered perfect. 
Likewise, they also correlate -- to different extents depending on the DIB -- with extinction \citep[e.g.][]{Herbig93}, neutral hydrogen, interstellar Na\,I\,D and Ca\,H\&K lines \citep[e.g.][]{Lan15,Friedman11}, 
and other molecules in the ISM \citep[e.g.\ CH, C$_2$, CN,][]{Weselak04,Baron15b}. In contrast, they are poorly correlated with the amount of H$_2$ \citep[e.g.][]{Lan15,Friedman11}.
Besides, different DIBs react in different ways to the UV radiation field \citep[e.g.][]{Krelowski92,Cami97,Cox06b,Vos11,Cordiner13,Elyajouri17}
and seem to show small scale spatial variations \citep{Wendt17,MonrealIbero15b,Cordiner13,vanLoon09} and different levels of clumpiness \citep{Kos17}.

An important piece of information is the behaviour of DIBs outside of the Milky Way \citep[see][for an excellent review of the subject]{Cordiner14}. 
Galaxies other than ours span a large diversity in terms of stellar, gas and dust content, mass, morphology, evolutionary status, environmental conditions, amongst others.
This entails a diversity of ISM properties, and thus an invaluable opportunity to test the existence and properties of DIBs under gas physical (e.g.\ radiation field, temperature, density) and chemical (e.g.\ metallicity, relative abundances, molecular content) conditions  and dust properties not found in the Milky Way.
However, since they are faint features, works aiming to study extragalactic DIBs are still scarce. 

The first DIB detected outside the Milky Way was that at $\lambda$4428, in the Magellanic Clouds \citep{Hutchings66,Blades79}. 
Nowadays, there are certainly many more DIB detections within these galaxies \citep{Ehrenfreund02,Cox06,Cox07,Welty06,Bailey15,vanLoon13}, reaching up to several tens of lines of sight for the DIBs at $\lambda$5780 and $\lambda$5797.
In both galaxies, DIBs seem weaker with respect to N(\ion{H}{i}) and \ebv\,  than in our Galaxy. The lower metallicity of the systems and, perhaps, the increased radiation field have been proposed as explanations \citep{Cox06,Welty06}.
Additionally, these same DIBs were also measured in a few lines of sight for the Andromeda galaxy \citep[\object{M\,31},][]{Cordiner11} and in the Triangulum galaxy \citep[\object{M\,33},][]{Cordiner08a}.
Outside of the Local Group, DIB detections are scant.
They have been found in relatively nearby galaxies that show high extinction \citep[e.g][]{Ritchey15,Heckman00}, towards bright supernovae \citep{DOdorico89,Thoene09,Cox08,Welty14,Sollerman05}, and damped Lyman-$\alpha$ systems \citep[e.g.][]{Junkkarinen04,Srianand13}.

As demonstrated in some of the examples mentioned above, highly multiplexed spectrographs offer the possibility of collecting in one shot information for many lines of sight. Integral field spectrographs (IFSs) constitute a special case of this kind of instrumentation where this information is recorded in an extended continuous field.
In \citet{MonrealIbero15}, we proposed the use of high-sensitivity IFSs as a tool to detect and map DIBs  outside the Local Group.
The experiment, conducted using the Multi Unit Spectroscopic Explorer (MUSE) commissioning data, led to the first determination of a DIB radial profile in a galaxy outside the Local Group. 
The experiment demonstrated the possibilities of using IFSs in extragalactic DIB research, but the strength of the DIB had to be determined only as average values in large areas (i.e.\ between 0.7 and 3.8~kpc$^2$).

Building on top of that experiment, we have initiated a search for DIBs in the galaxies observed within the MUSE Guaranteed Time Observations. Here, we present the results for the \object{Antennae Galaxy} (\object{NGC\,4038}/\object{NGC\,4039}, \object{Arp\,244}). 
This system is one of the most spectacular examples of gas-rich major mergers as well as being the closest (distance=22$\pm$3~Mpc, \citealt{Schweizer08}).
The system has a total infrared luminosity of $\log(L_{\mathrm{ir}} /L_\odot) = 10.86$ \citep{Sanders03}
approaching that of a luminous infrared galaxy, and is subject of heavy and variable attenuation \citep{Brandl05}. It therefore presents a good target for the search of extragalactic DIBs.

The paper is structured as follows:
Sect.~\ref{sec:data} summarises the data collection and reduction;
Sect.~\ref{sec:extract} contains a description of our methods to extract the information for the DIBs and other used observables. 
Results are presented and discussed in Sect.~\ref{sec:disc}.
Our main conclusions are summarised in Sect.~\ref{sec:concl}.

\begin{figure}[th]
\centering
\includegraphics[angle=0,width=0.48\textwidth, clip=, viewport=30 60 540 500,]{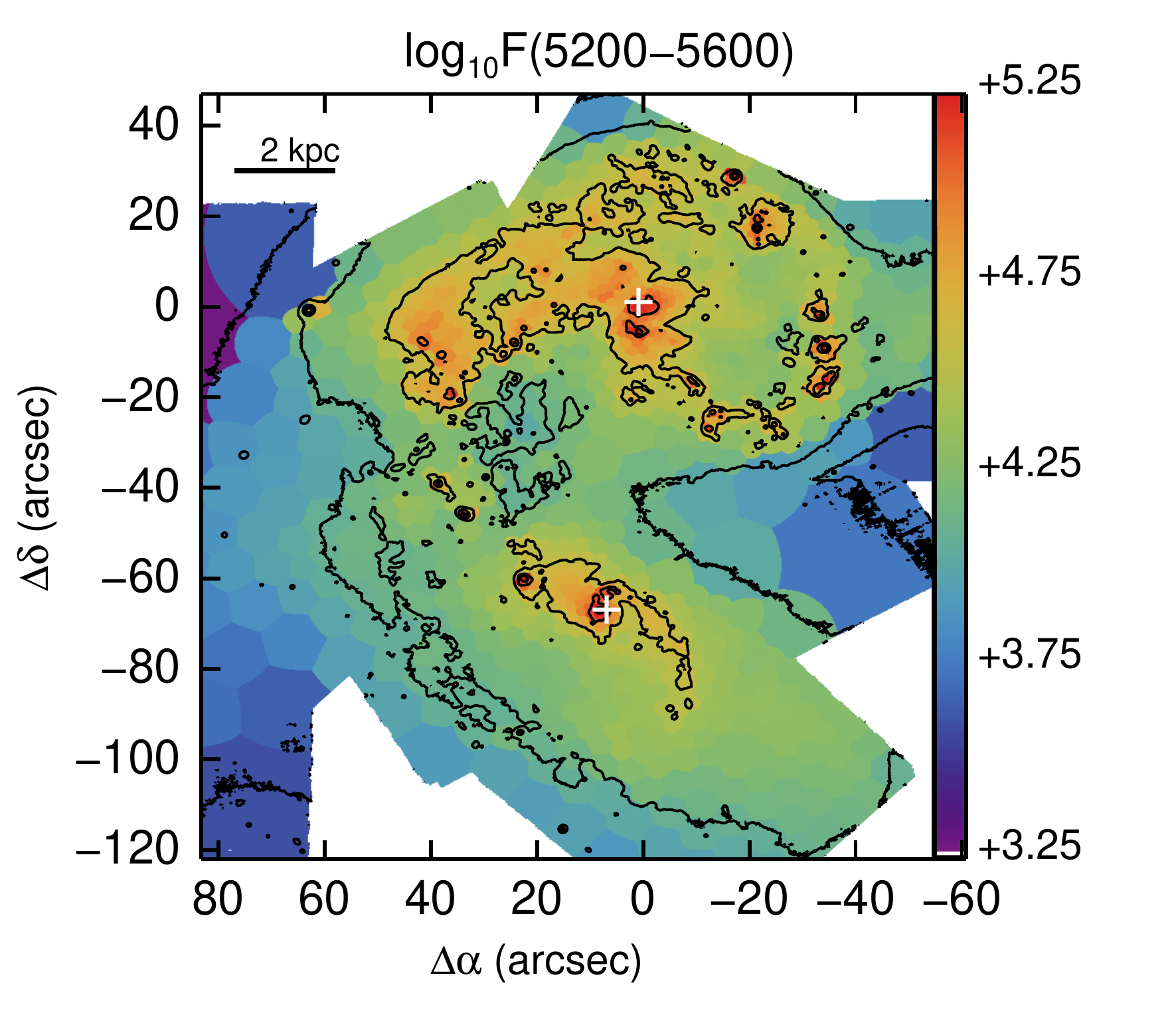}
  \caption[]{Line-free continuum as derived from \texttt{STARLIGHT} showing the mapped area with the tessellated pattern used to extract the high signal-to-noise ratio spectra (i.e.\ the tiles).
The reconstructed white-light image is overplotted as reference with  contours in logarithmic stretching of 0.5 dex steps.
The origin of coordinates here and through the paper was put at the position of the northern nucleus. North is up, east is to the left.}
\label{apuntado}
\end{figure}

\begin{figure*}
\centering
\includegraphics[angle=0,width=0.49\textwidth, clip=, viewport=0 0 780 480,]{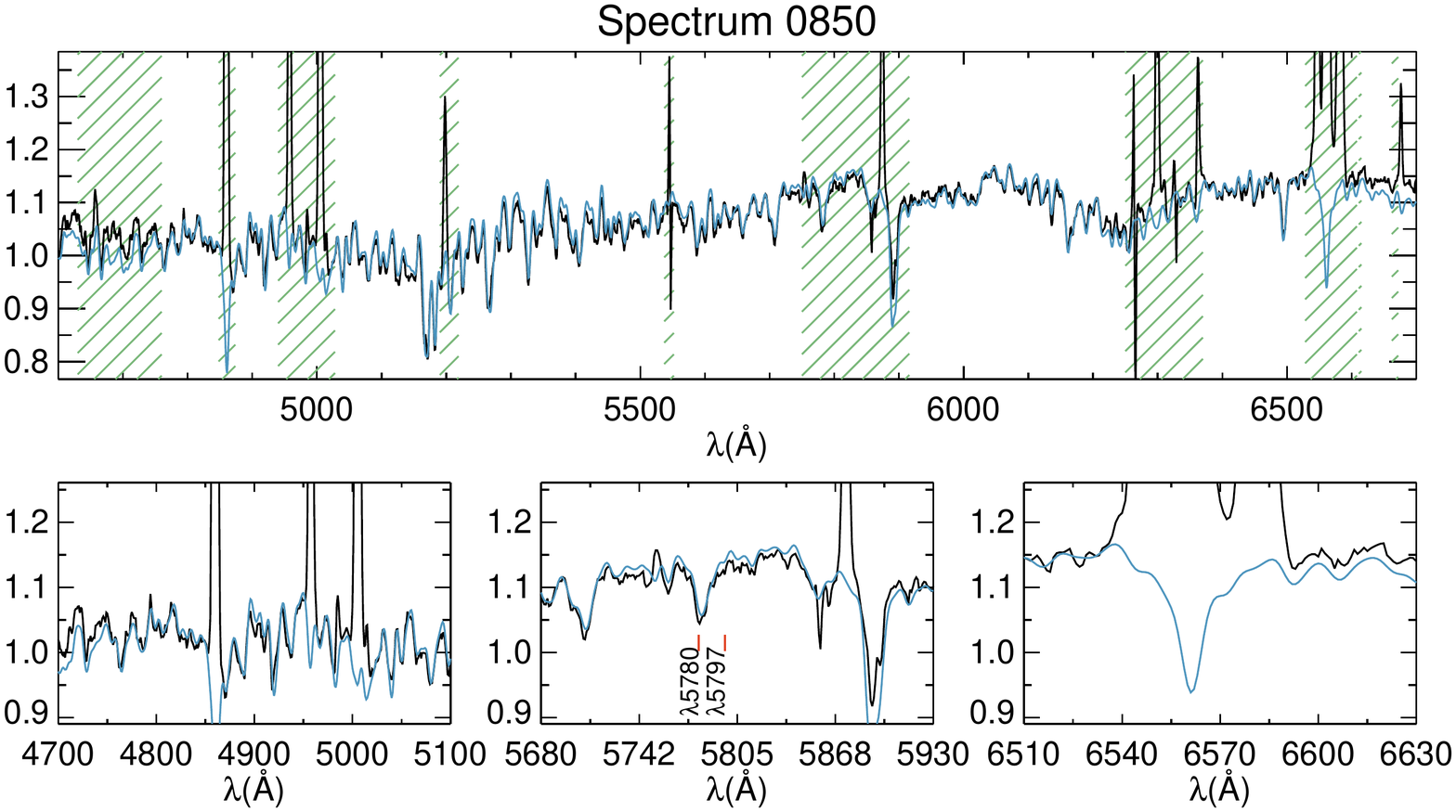}
\includegraphics[angle=0,width=0.49\textwidth, clip=, viewport=0 0 780 480,]{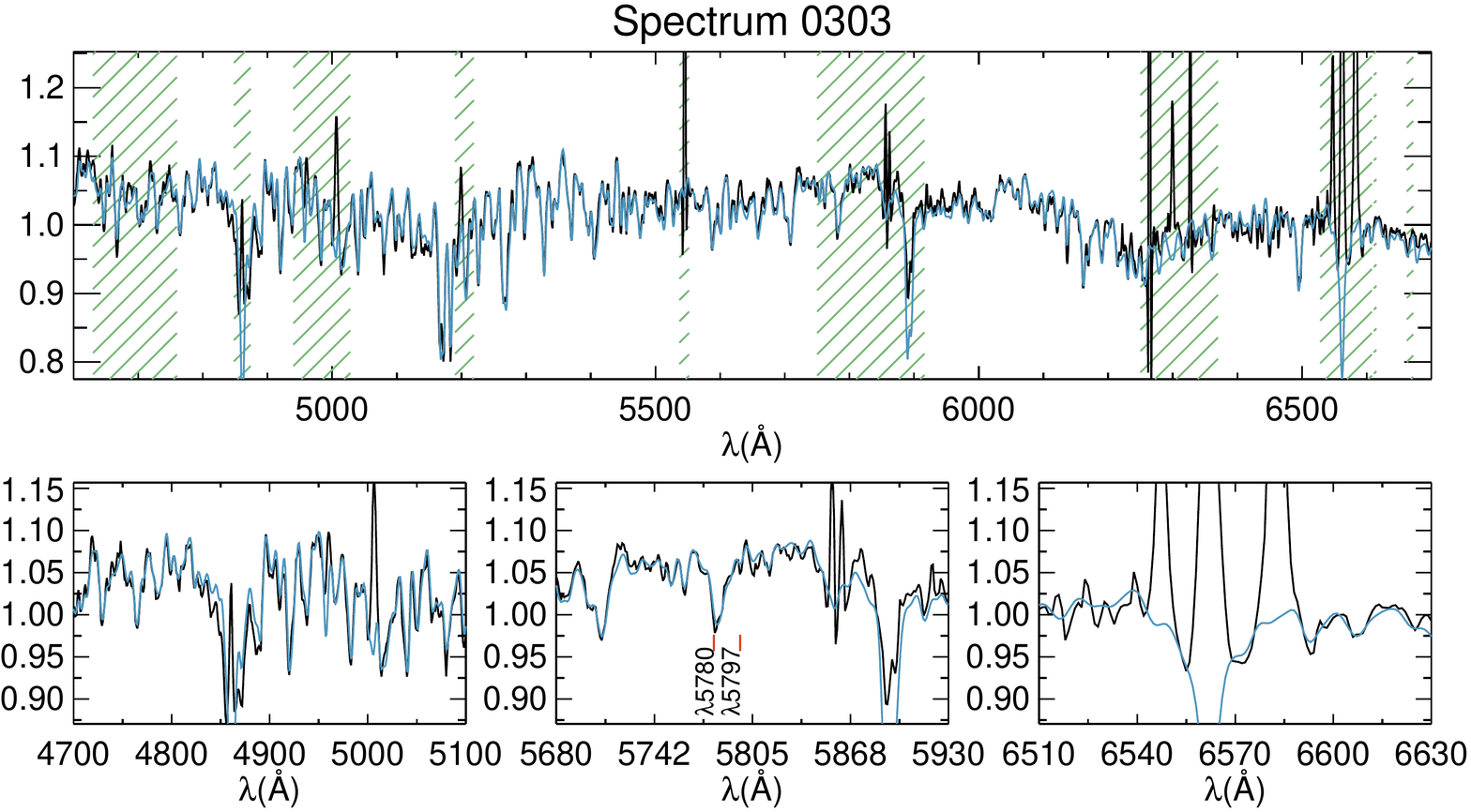}

\includegraphics[angle=0,width=0.49\textwidth, clip=, viewport=0 0 780 480,]{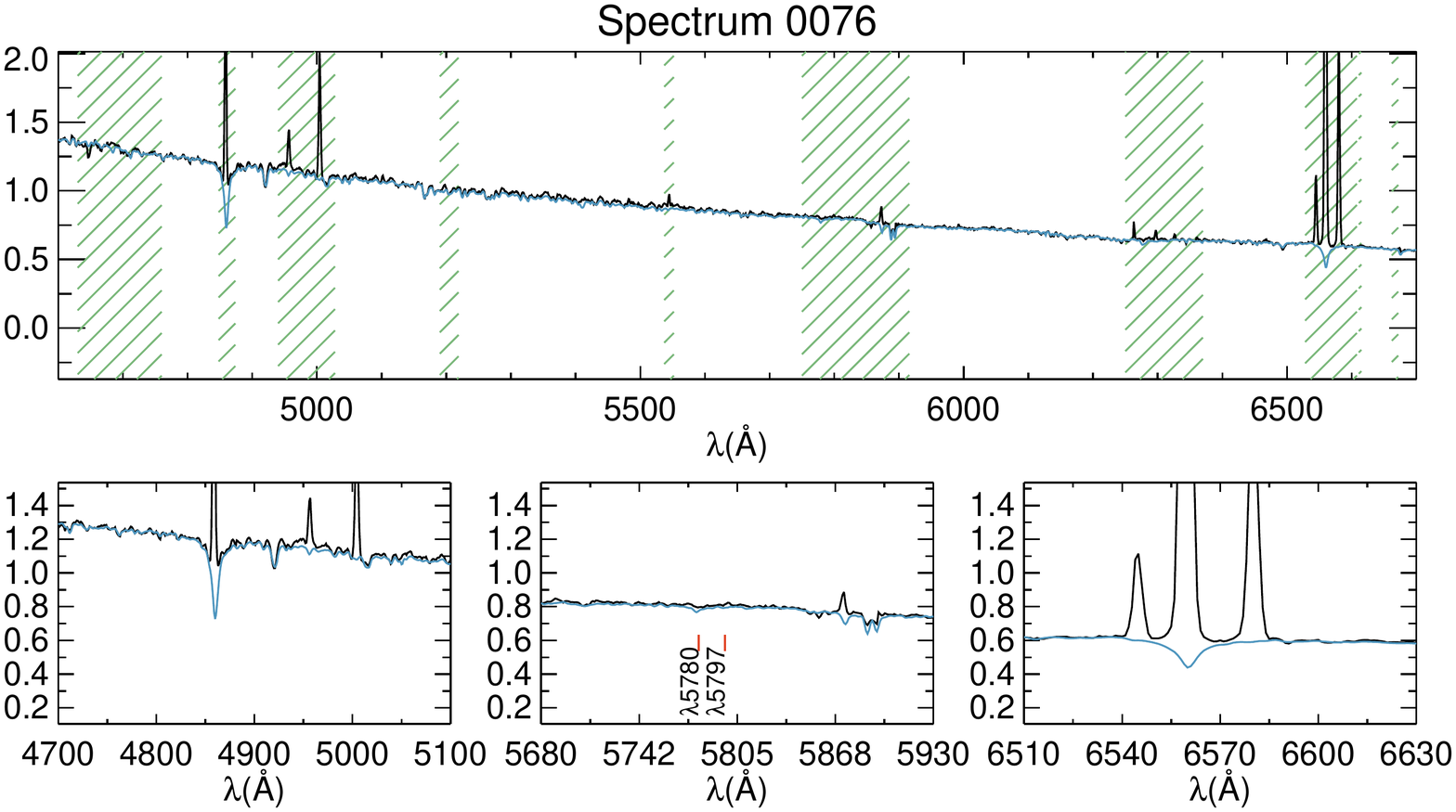}
\includegraphics[angle=0,width=0.49\textwidth, clip=, viewport=0 0 780 480,]{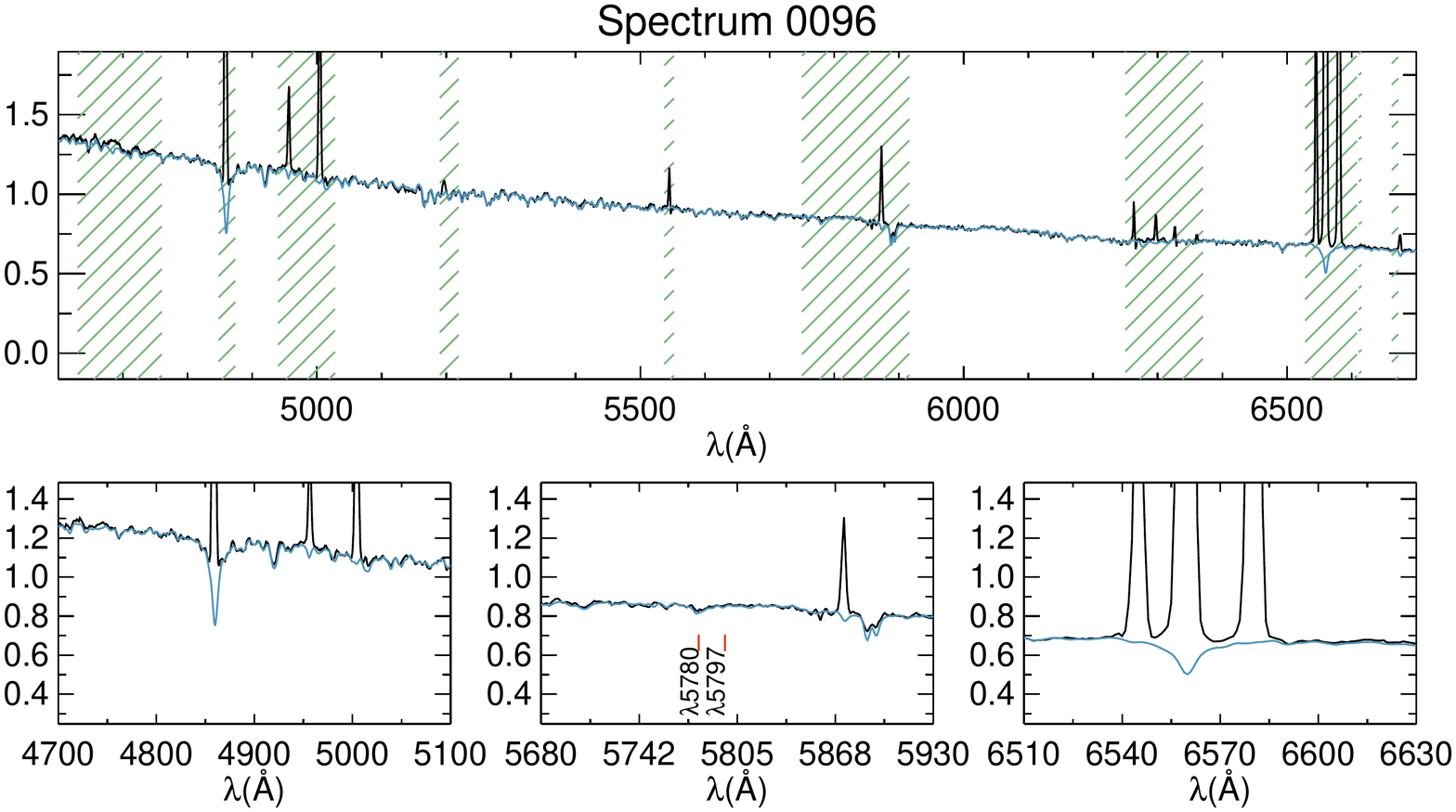}

\caption[]{Examples modelling stellar populations.
Each panel displays the complete modelled spectral range in the upper row and three zooms in key spectral regions in the lower row. The observed spectrum is in black and the modelled stellar spectral energy distribution in light blue. Masked spectral ranges are marked using bands with green stripes in diagonal.
The positions of the DIBs at $\lambda$5780 and $\lambda$5797 are marked in the central zoom of each panel with red ticks.
\emph{Upper row:} Two red spectra without DIB detection. 
\emph{Lower row:} Two blue spectra without DIB detection.
Reported wavelengths here and throught the paper are in rest-frame.
\label{figfitstellarpopA}
}
\end{figure*}

\begin{figure*}
\centering
\includegraphics[angle=0,width=0.49\textwidth, clip=, viewport=0 0 780 480,]{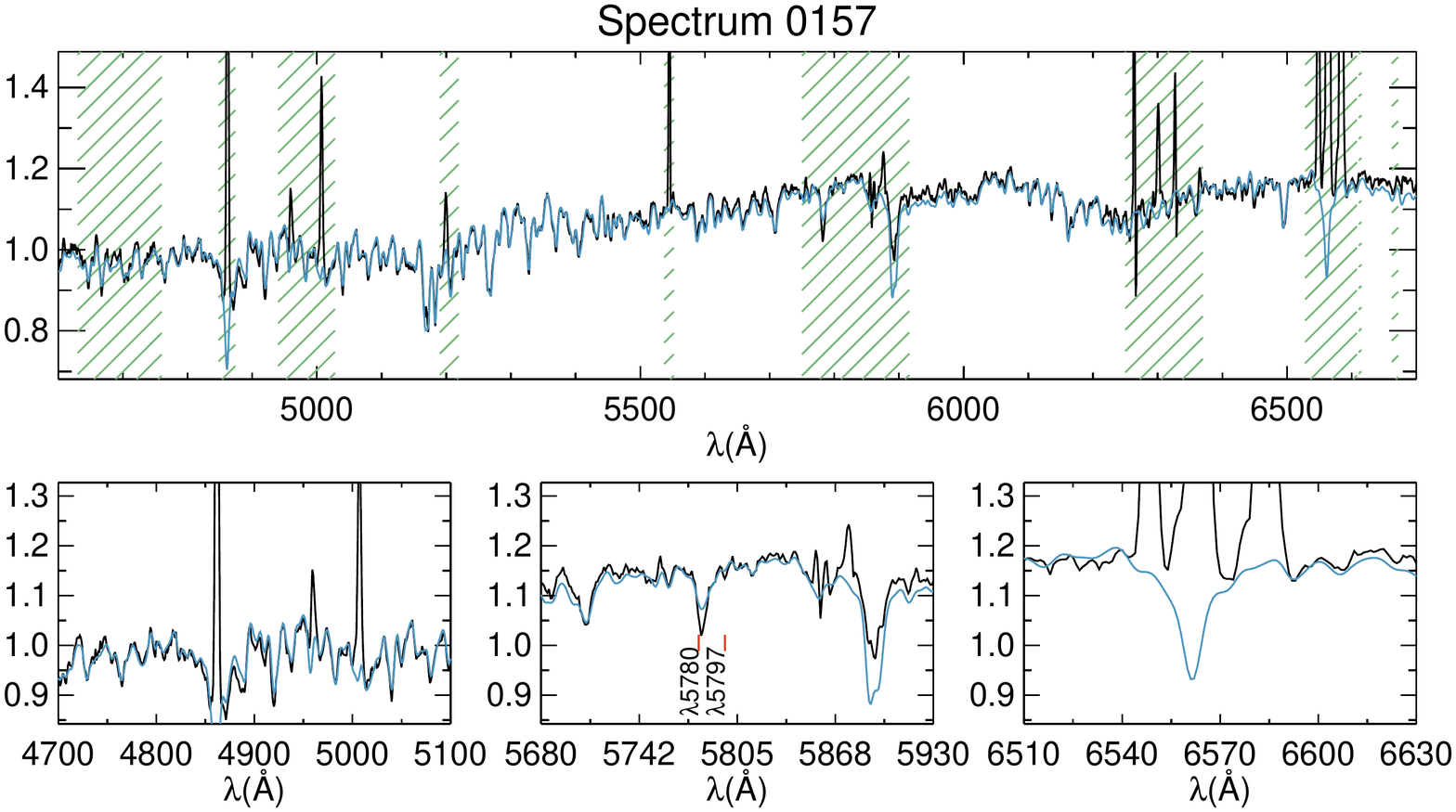}
\includegraphics[angle=0,width=0.49\textwidth, clip=, viewport=0 0 780 480,]{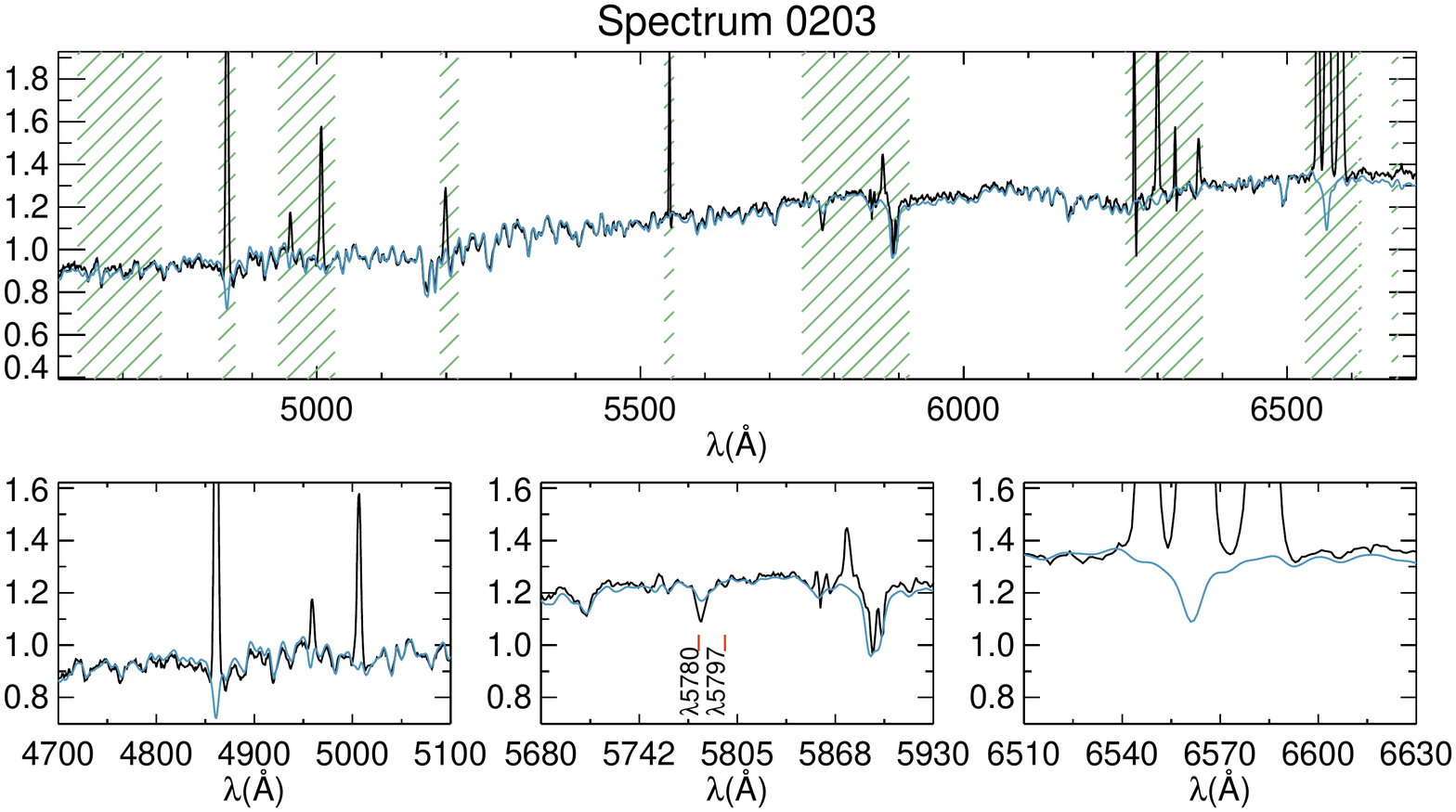}

\includegraphics[angle=0,width=0.49\textwidth, clip=, viewport=0 0 780 480,]{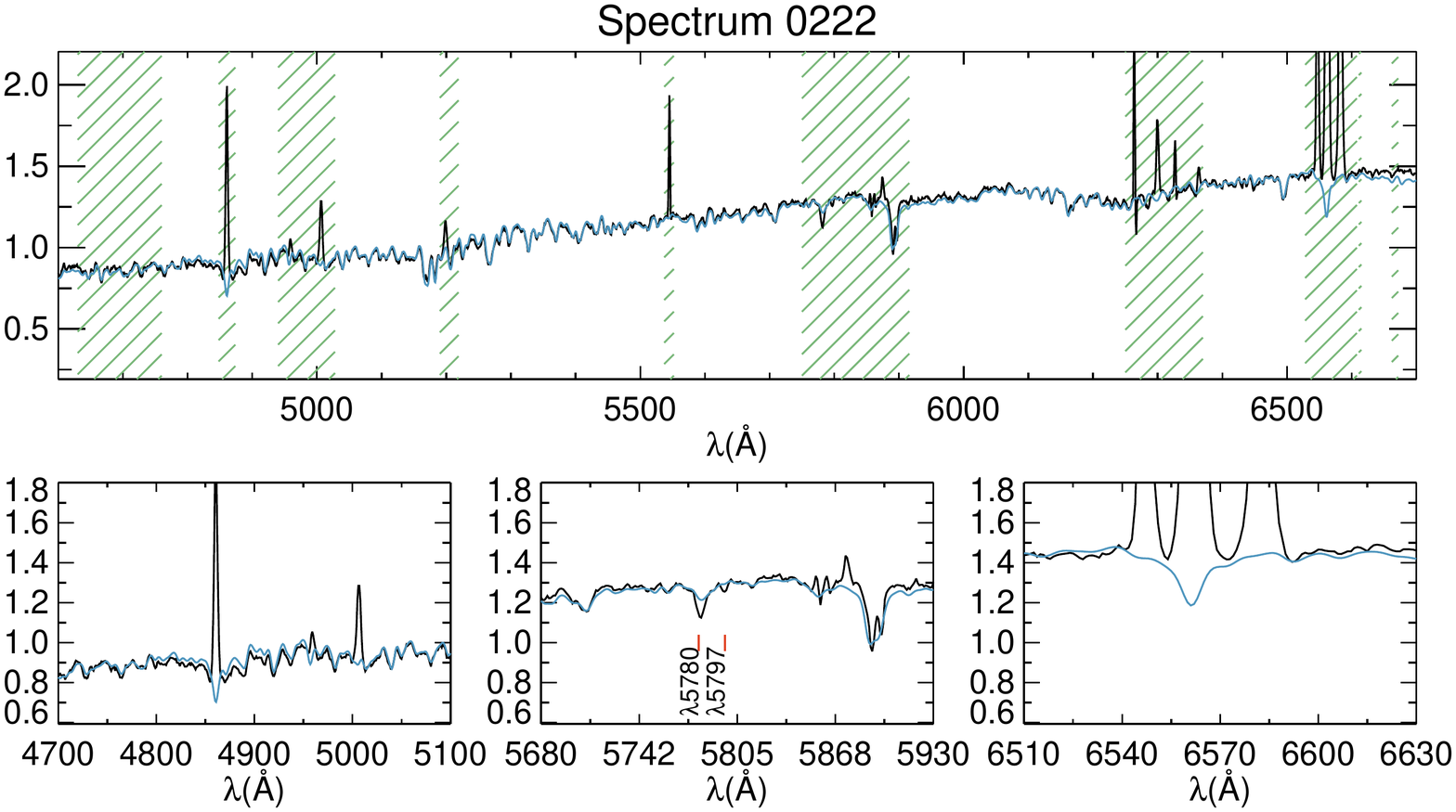}
\includegraphics[angle=0,width=0.49\textwidth, clip=, viewport=0 0 780 480,]{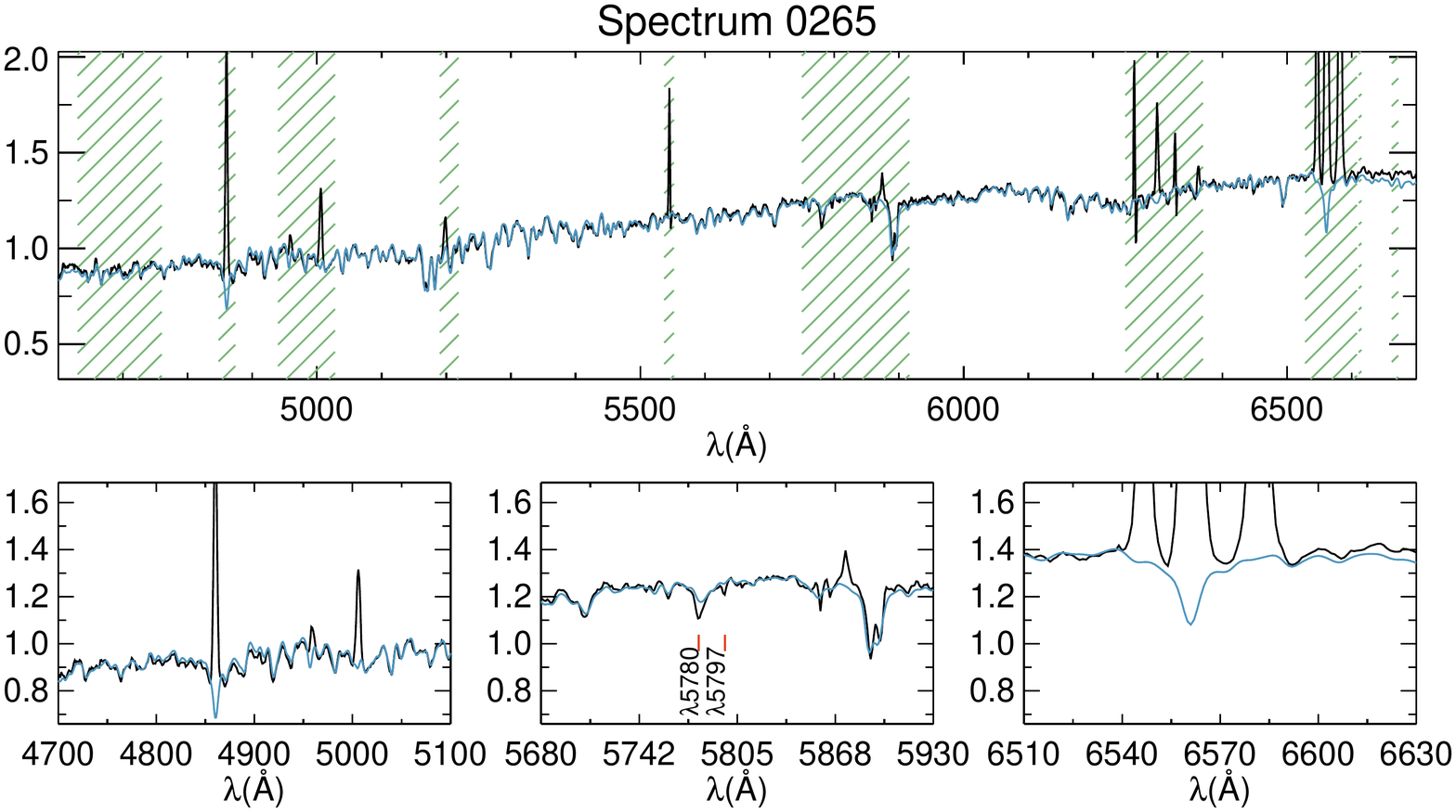}
\caption[]{Same as Fig.~\ref{figfitstellarpopA} but for spectra with DIB detection.
\emph{Upper row:} Two spectra with a clear residual only at the position of  the DIB at $\lambda$5780. 
\emph{Lower row:} Two spectra with apparent residuals at the position of both, the DIB at $\lambda$5780 and that at  $\lambda$5797.
}
\label{figfitstellarpopB}
\end{figure*}

\section{The data}\label{sec:data}
\defcitealias{WMV+17}{Paper~I} 
The dataset we use in this paper has already been presented in great detail in
\citet[][hereafter Paper~I]{WMV+17}. Here, we give a short summary of its
properties and the processing involved.

The central merger of the NGC\,4038/39 system was observed in April/May 2015
and February and May 2016 with MUSE on VLT UT4 \citep{Bacon12}. The wide-field
mode in extended configuration was used, which yields a sampling of 0\farcs2
over a field of about $1\sq\arcmin$, with a wavelength range of 4600 to
9350\,\AA, sampled at about 1.25\,\AA\,pixel$^{-1}$.
As explained in \citetalias{WMV+17}, we reduced the data using the public MUSE
pipeline \citep[v.~1.6;][Weilbacher et al.\ in prep.]{2014ASPC..485..451W}, using
standard calibrations, including bias subtraction, flat-fielding and slice
tracing, wavelength calibration, twilight sky correction, atmospheric
refraction correction, and sky subtraction (using external sky fields) as well
as correction to barycentric velocities. All exposures of the central field
were then combined, using the linear full width at half maximum (FWHM) weighting scheme built into the
pipeline, to optimise the final seeing. We minimised contributions from exposures of lower sky transparency by resetting
their FWHM keywords to high values. Most of the exposures were 1350\,s long,
but we replaced one small section that was saturated at \ha\ by two 100\,s
exposures. The final output datacube has $950\times973\times4045$
voxels\footnote{In this paper, we only used the linearly sampled version of the
  cube.}, with an effective seeing of about 0\farcs6 FWHM at 7000\,\AA. Due to
the irregular layout of the observations, some of the spatial regions in the
cube remain empty. A reconstructed continuum image of the mapped area used in this work is displayed as contours in Fig.~\ref{apuntado}.
Please see \citetalias{WMV+17} for other information about the data.

\section{Extracting information}\label{sec:extract}

\subsection{Tile definition and modelling of the underlying stellar population emission \label{secmodelling}}

The cube was tessellated by grouping and co-adding the spectra using the Voronoi binning technique \citep{Cappellari03} to produce data of $S/N\sim 250$ in the mostly emission-line free wavelength range 5600\dots6500\,\AA.
Then, the stellar spectral energy distribution in each tile was modelled and subtracted. This is necessary to properly derive the fluxes for the Balmer emission lines, and crucial to identify the faint ISM features in absorption.
To do so, we used the \texttt{STARLIGHT}\footnote{\texttt{http://starlight.ufsc.br/index.php?section=1}} spectral synthesis code \citep{cid05,cid09} that reproduces a given observed spectrum by selecting a linear combination of a sub-set of $N_{\ast}$ spectral components from a pre-defined set of base spectra.
Our aim was not to in detail determine the characteristics of the stellar population but to recover the information for the ISM component, both in emission and absorption.
In our particular case, we modelled the spectral range from 4570\,\AA\, to 6800\,\AA\, (rest-frame). 
We used a set of simple stellar populations of three different metallicities (Z=0.004, 0.02, and 0.05) from \citet{bru03} which are based on the Padova 2000 evolutionary tracks \citep{gir00} and assumed a Salpeter initial mass function between 0.1 and 100~M$_\odot$.
For each metallicity, there was a total of 15 spectra with ages ranging from 1 Myr to 13 Gyr.
Critical spectral ranges corresponding to ionised gas emission lines, Wolf-Rayet star emission, possible ISM absorption features, and sky substraction residuals were masked out for the fit.
Finally, we assumed a common extinction for all the base spectra, and modelled it as a uniform dust screen with the extinction law by \citet{car89}. 

All the main outputs for each tile were then reformatted in order to create four files with the total, modelled stellar emission, residuals, and ratio of total to stellar emission.
Additionally, the information associated to auxiliary
properties such as stellar extinction in the $V$ band, the reduced $\chi^2$ and the absolute
deviation of the fit (in \%) were reformatted and saved as 2D FITS images suitable for manipulation with standard astronomical software. 

 use both map and image to refer to this kind of file.

The  reduced $\chi^2$ was relatively uniform all over the covered area with a mean ($\pm$standard deviation) $\chi^2$ of 2.23($\pm$0.43). The same applies to the absolute
deviation of the fit, which was of $\sim$0.4\%. To have an idea of the tessellation pattern for the mapped area, a line-free continuum is presented in Fig.~\ref{apuntado}.
Some representative examples of the modelled spectra are found in Figs.~\ref{figfitstellarpopA} and \ref{figfitstellarpopB}. Each panel displays the complete modelled spectral range in the upper row and three zooms in key spectral regions in the lower row. The positions of the DIBs at $\lambda$5780 and $\lambda$5797 are marked in the central enlargement of each panel.

Spectra in Fig.~\ref{figfitstellarpopA} are examples with no apparent detection of the DIBs at $\lambda$5780 and $\lambda$5797. These spectra could be both: relatively red (upper row), or relatively blue (lower row).
Spectra in Fig.~\ref{figfitstellarpopB} are examples with DIB detections. These are all spectral with a red continuum. In the upper row, only the residual in the modelling of the spectra at $\lambda$5780 is apparent, at first sight. In the lower row both residuals, at $\lambda$5780 and $\lambda$5797, are clearly seen.

\begin{figure*}[!htb]
\centering
\includegraphics[angle=0,width=0.24\textwidth, clip=, viewport=60 50 390 320,]{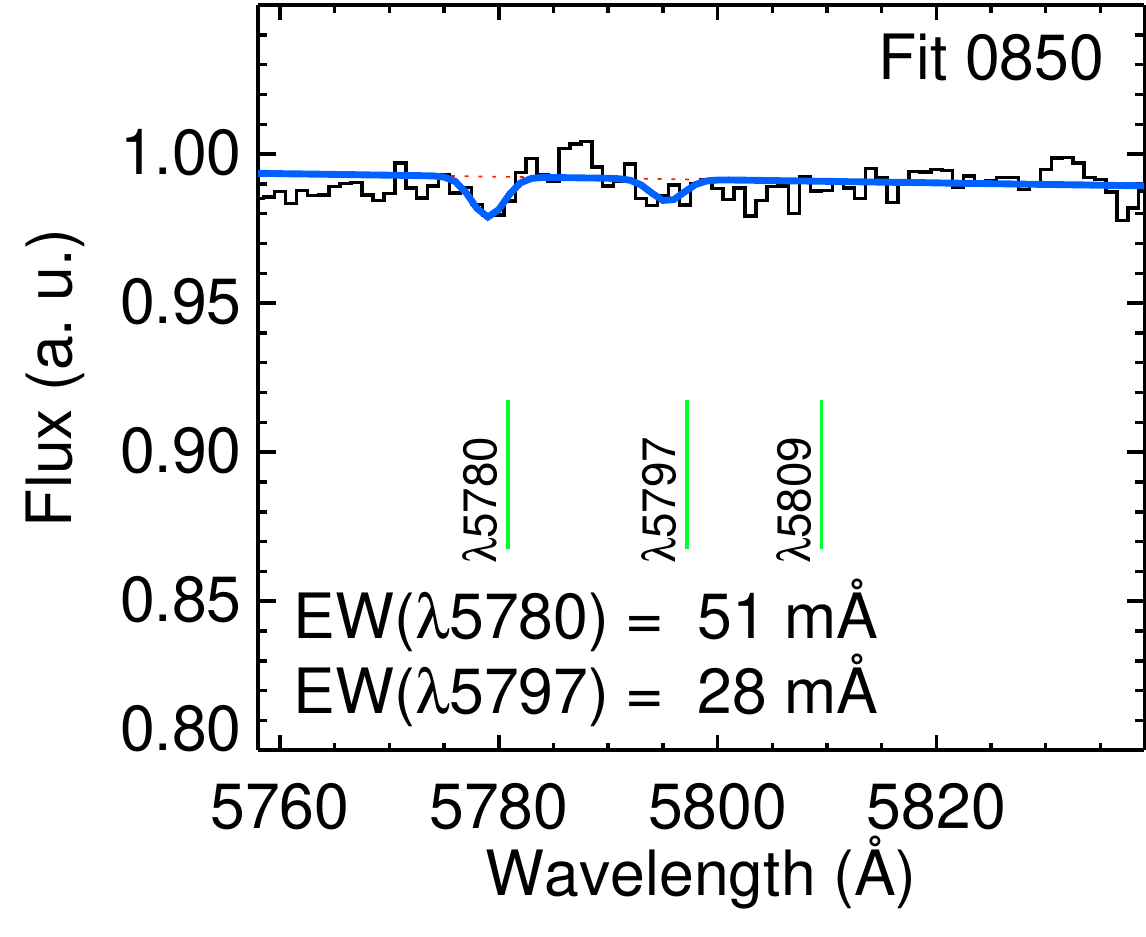}
\includegraphics[angle=0,width=0.24\textwidth, clip=, viewport=60 50 390 320,]{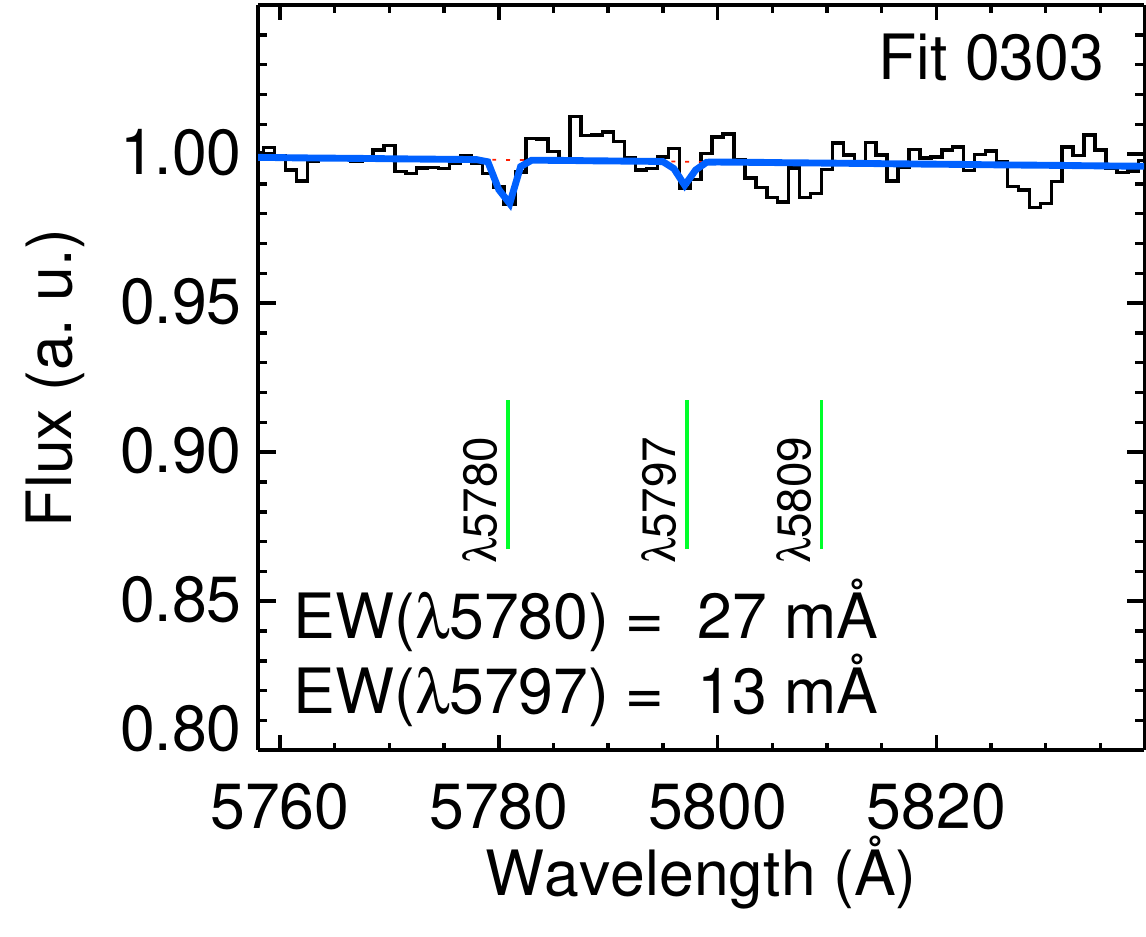}
\includegraphics[angle=0,width=0.24\textwidth, clip=, viewport=60 50 390 320,]{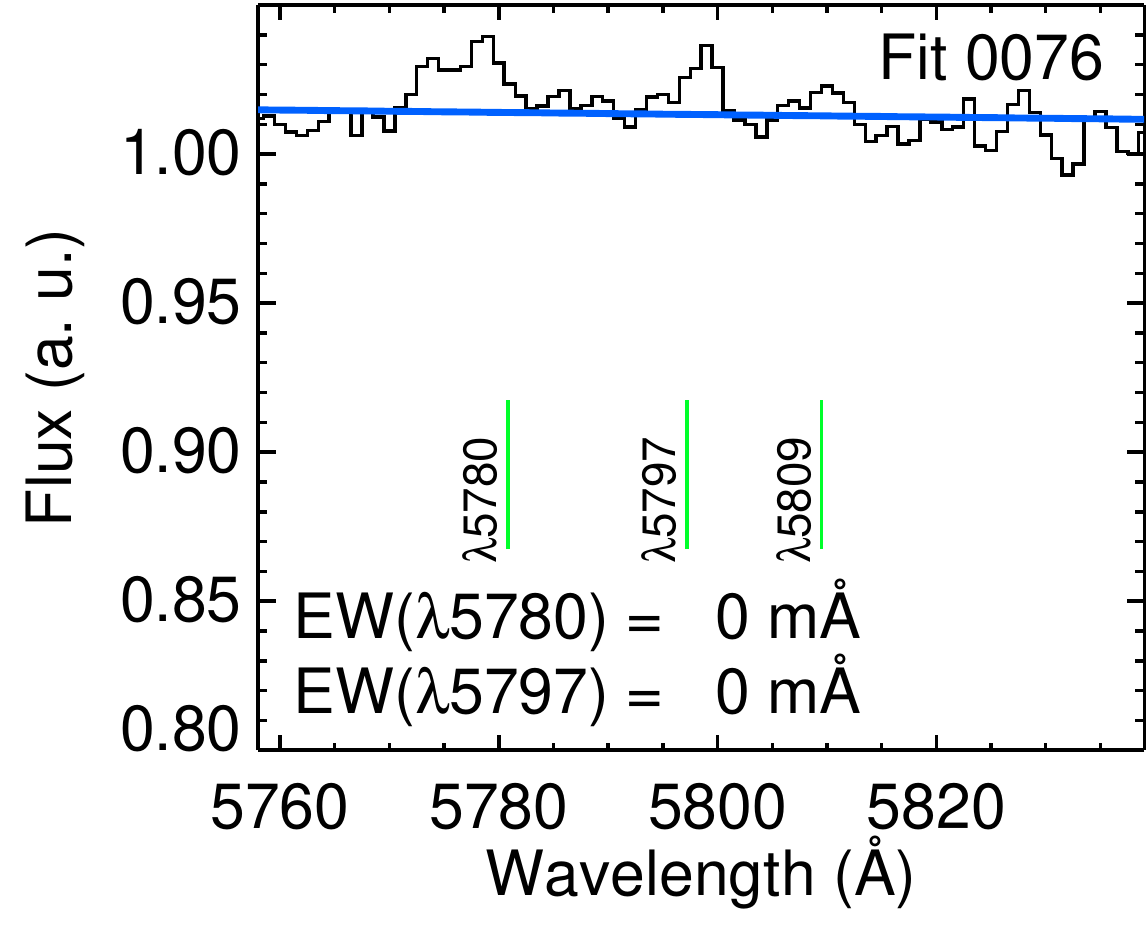}
\includegraphics[angle=0,width=0.24\textwidth, clip=, viewport=60 50 390 320,]{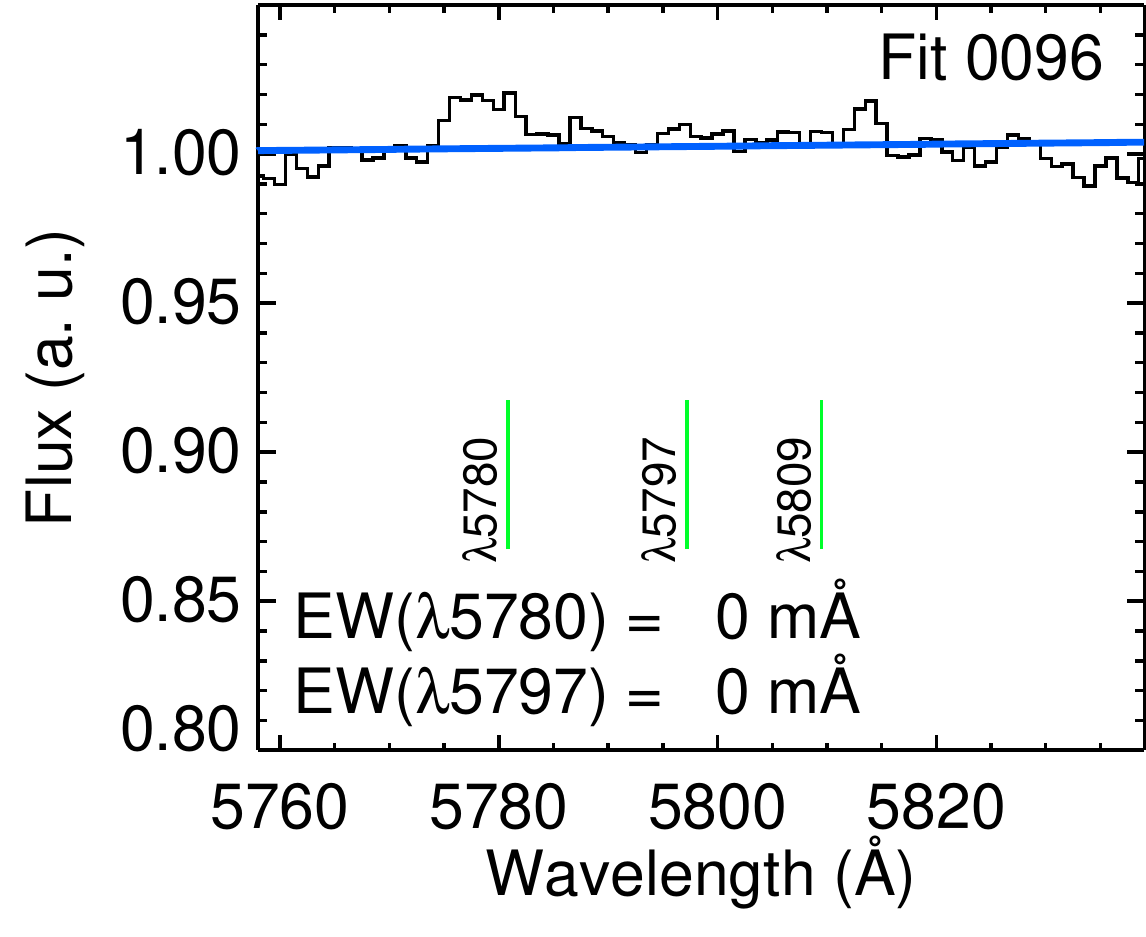}

\includegraphics[angle=0,width=0.24\textwidth, clip=, viewport=60 50 390 320,]{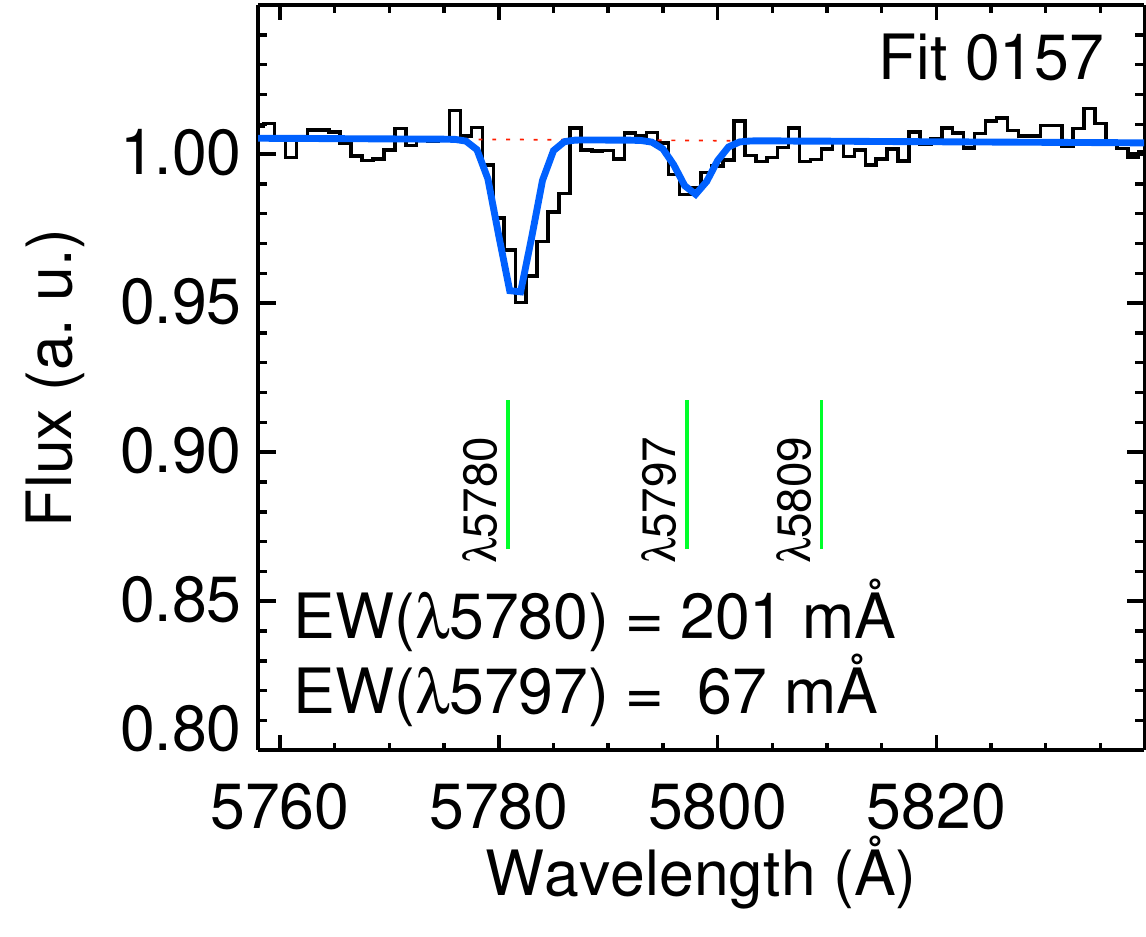}
\includegraphics[angle=0,width=0.24\textwidth, clip=, viewport=60 50 390 320,]{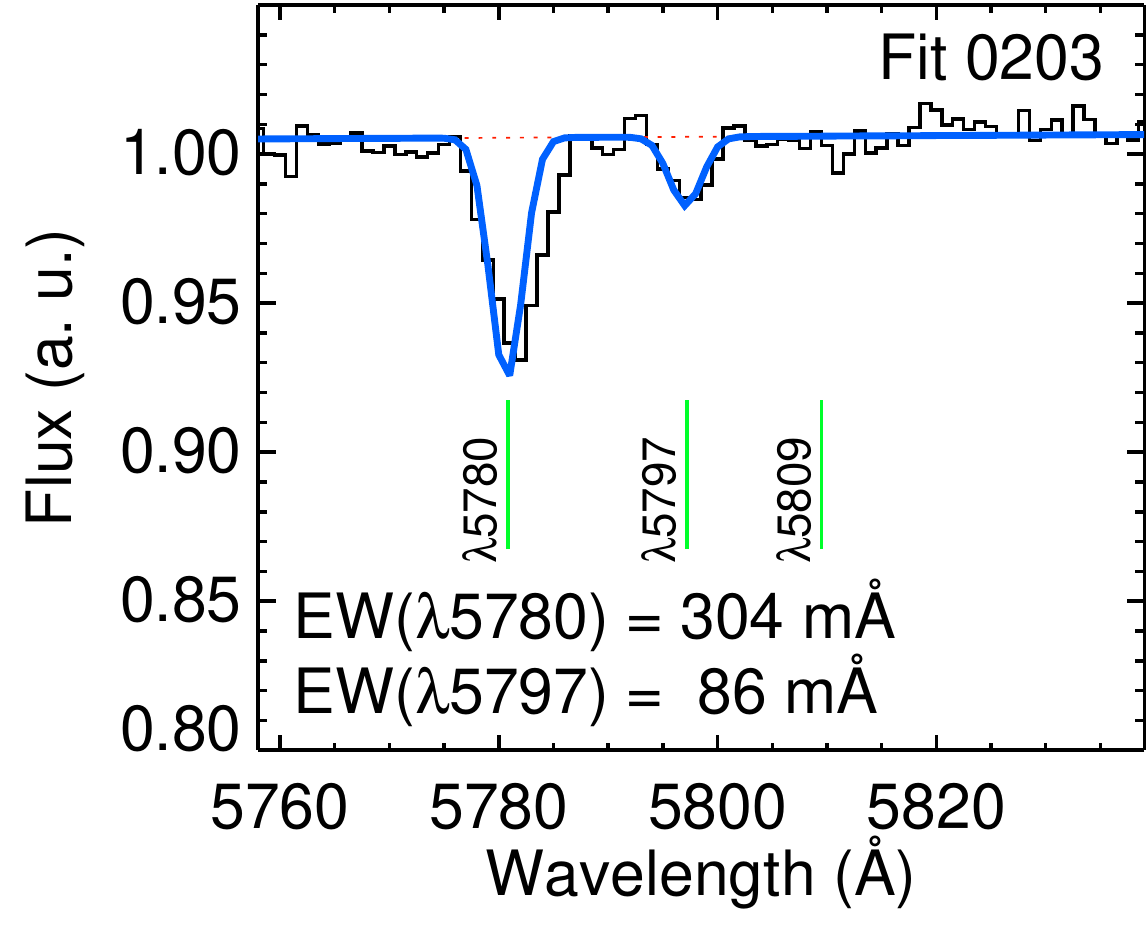}
\includegraphics[angle=0,width=0.24\textwidth, clip=, viewport=60 50 390 320,]{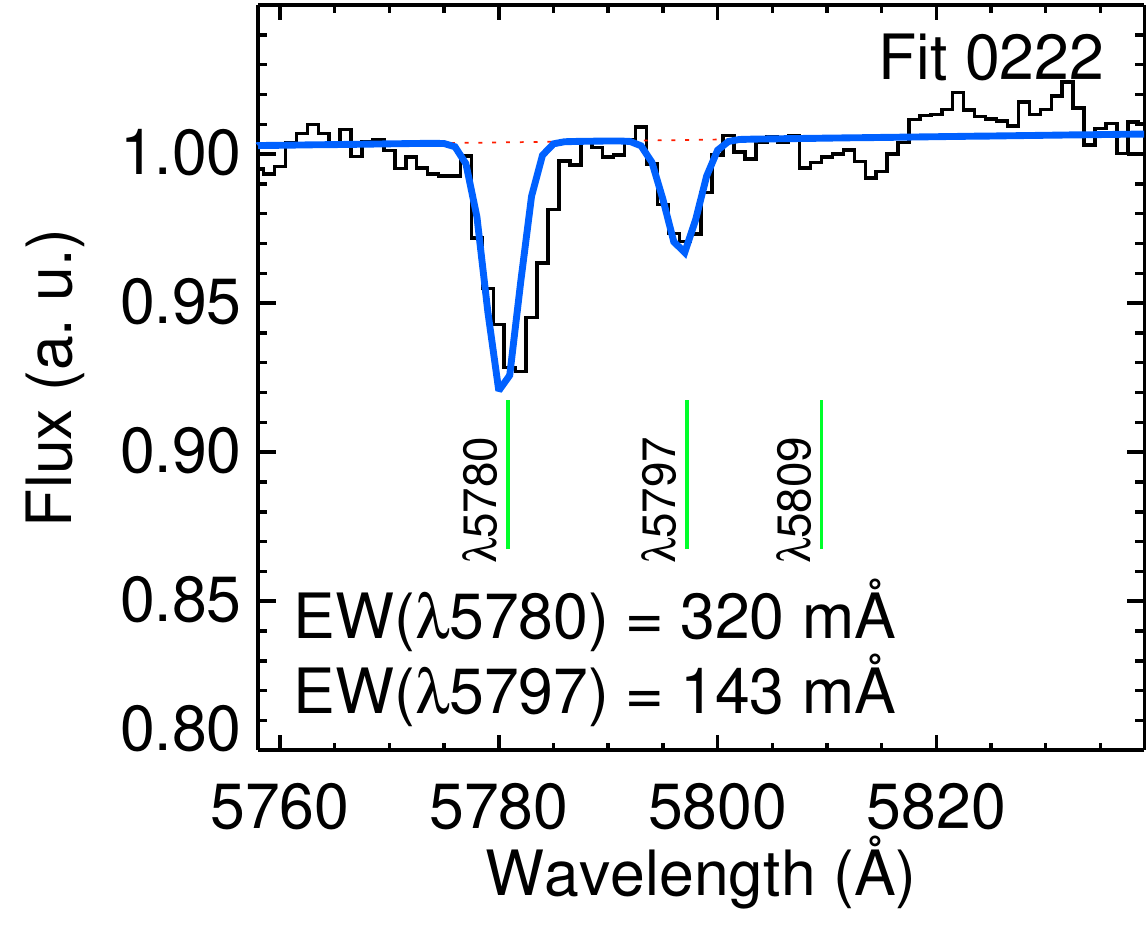}
\includegraphics[angle=0,width=0.24\textwidth, clip=, viewport=60 50 390 320,]{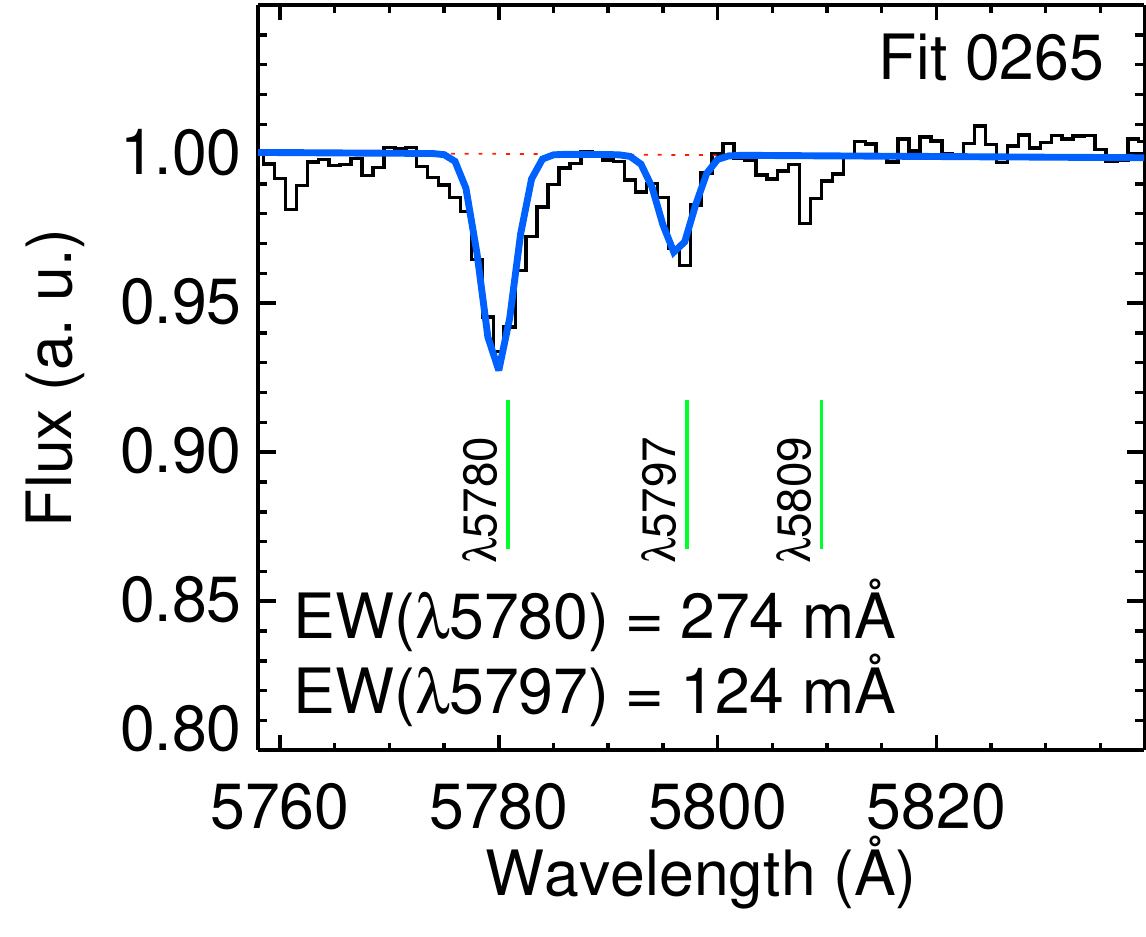}
\caption[]{DIB fit for the spectra
presented in Fig.~\ref{figfitstellarpopA} (\emph{upper row}) and Fig.~\ref{figfitstellarpopB} 
(\emph{lower row}). The ratio of observed  to modelled spectra are plotted in black while the total fits are in blue. Red dashed lines mark the fitted continua.
}
\label{exfits}
\end{figure*}

\begin{figure}[!htb]
\centering
\includegraphics[angle=0,width=0.22\textwidth, clip=, viewport=10 20 450 450,]{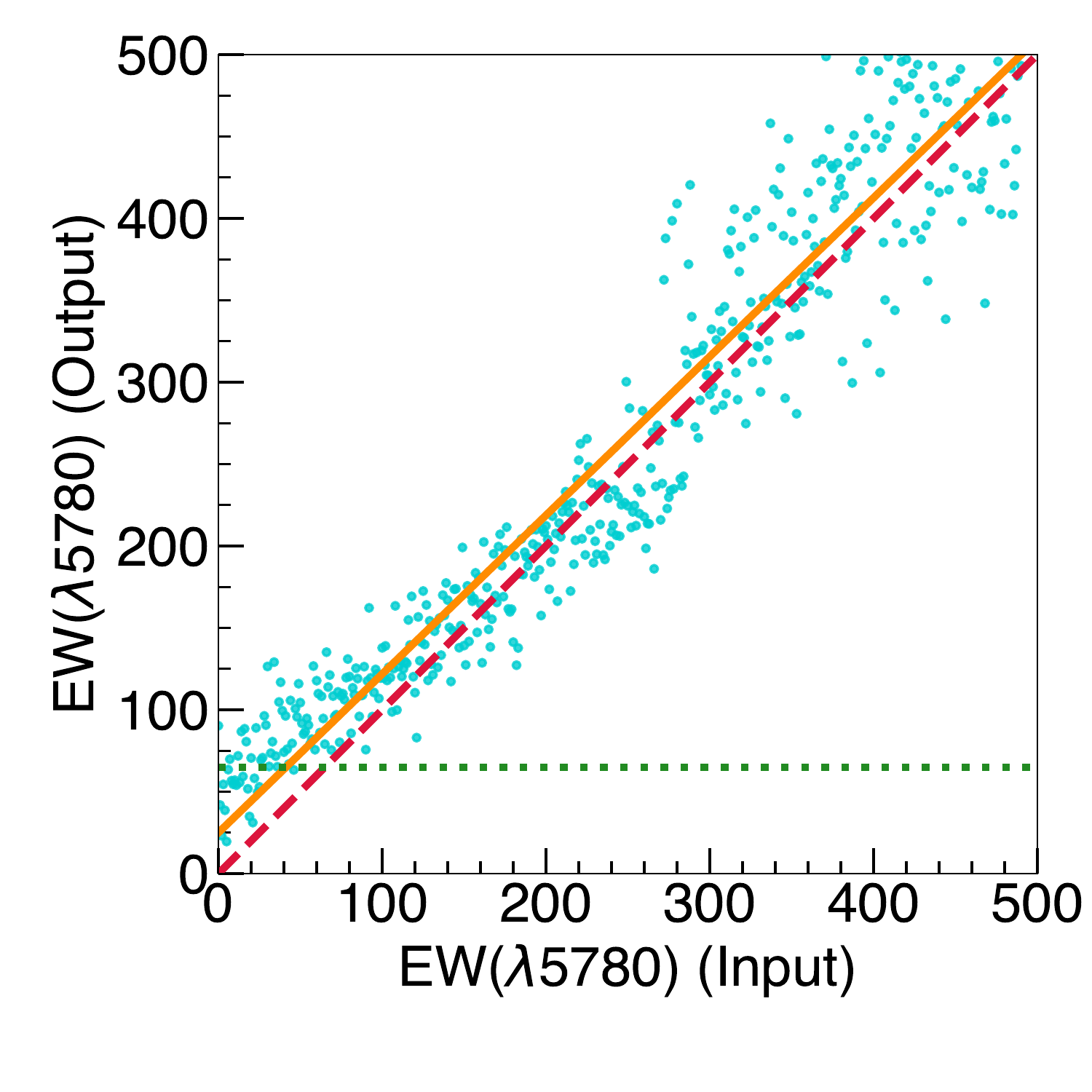}
\includegraphics[angle=0,width=0.22\textwidth, clip=, viewport=10 20 450 450,]{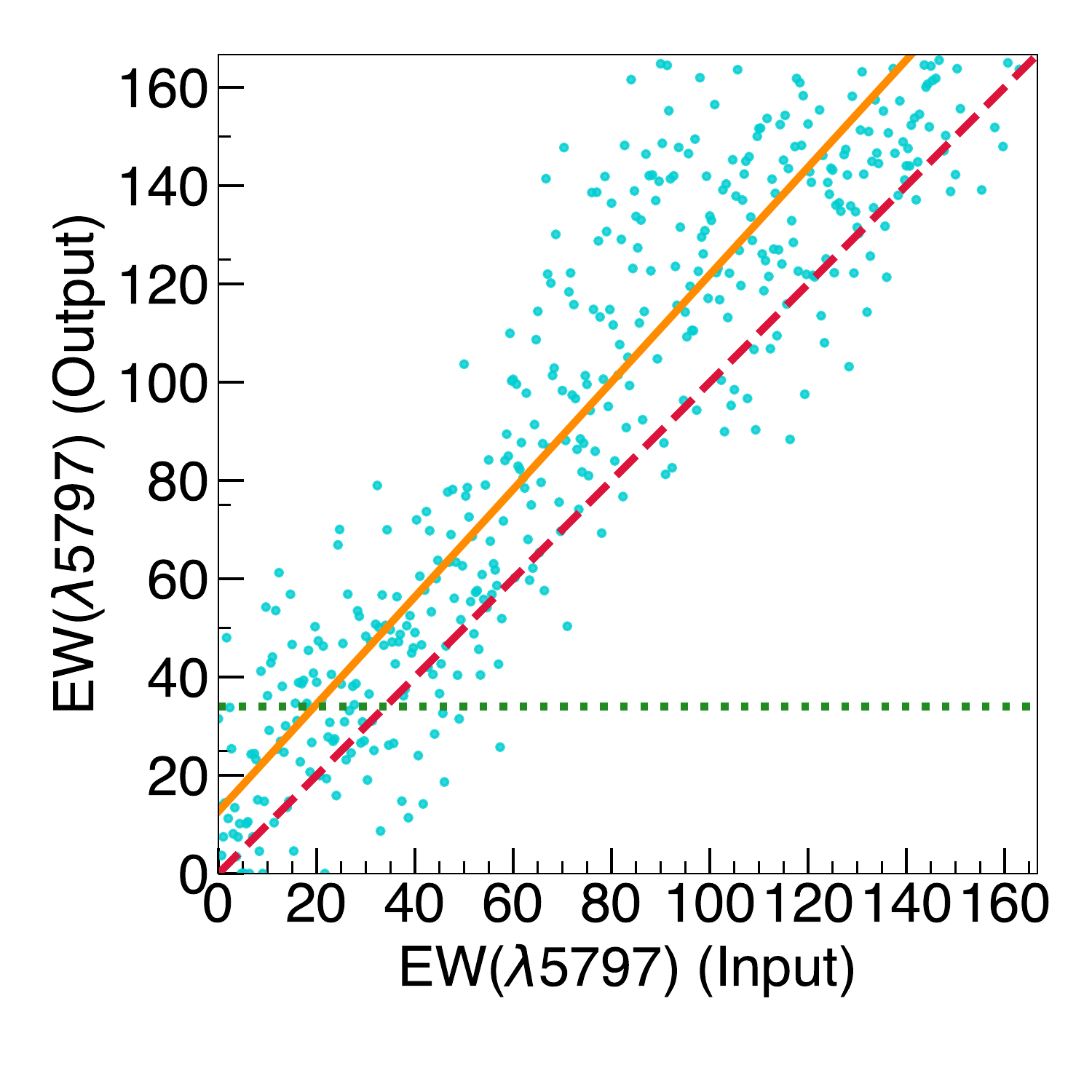}
\caption[Results for mock data]{Results for mock data. Measured equivalent width in m\AA\, as a function of the input in the mock data (\emph{left:} EW($\lambda$5780); \emph{right:} EW($\lambda$5797)).
The continuous orange line shows the one-degree polynomial fit while the red dashed line shows the one-to-one relationship. Our adopted  limit for DIB detection is marked with a dotted green line.
}
\label{figcomapmockspec}
\end{figure}

\subsection{Spectral fitting \label{secfit}}

We fitted the different spectral features needed
for our analysis using the IDL-based routine \texttt{mpfitexpr} \citep{Markwardt09}. 
Both emission and absorption features were fitted to Gaussian functions.

The fitting to the strong emission lines was done in the residuals (\emph{''observed - modelled''}) spectra. The procedure is relatively standard and does not pose a particular challenge for the goals of this work. Therefore, we refer to \citet{MonrealIbero10a} for the details about it.

The fitting of the faint absorption features is, in turn, quite challenging. This was done in the ratio (\emph{''observed / modelled''}) spectra.
In this paper, we have focussed on the absorption features at $\lambda$5780 and $\lambda$5797.
Nonetheless, we inspected the spectra for the presence of other (stronger) DIBs and found evidences of, at least, the DIB at $\lambda$6284 in many of them.
This DIB requires a more delicate correction of the telluric absorption and \oi$\lambda\lambda$6300,6364 emission lines than the one currently applied to this set of data. Therefore, we decided to postpone any experiment with this feature for future studies.

We reproduced the DIBs at $\lambda$5780 and $\lambda$5797 as two Gaussian functions fixing the relative wavelength between them.
At the resolution of the MUSE data, these features are a blend of several DIBs. We created a spectrum using the DIB compilation by \citet{Jenniskens94}, which was degraded and re-sampled to mimic the characteristics of the MUSE data. We then measured the relative central wavelengths there.
We allowed for absorption features only,
and assumed the same line width for both features.
The shape of these features is dominated by the instrumental spectral resolution of the MUSE data and thus, using Gaussian functions is a reasonable assumption.
We used the results for the fit of the \hb\, emission line in velocity and line width to establish the initial guess for the central wavelength and width of the absorption features. 
Additionally, we included a one-degree
polynomial to model the residuals in the stellar continuum.

We display some examples of fits in Fig.~\ref{exfits}. The upper row contains the fits for spectra in Fig.~\ref{figfitstellarpopA}, consistent with no DIB detection (see Sect.~\ref{secbias}).
Still, one can see a possible detection of the DIB at $\lambda$5780 in the spectrum 850 just below the significance level.
The lower row has the fits for the spectra in Fig.~\ref{figfitstellarpopB}.
A feature in absorption at the position of the 
DIB at $\lambda$5797 is clearly seen in all four cases, including spectra 157 and 203.
Likewise, the two examples with strongest DIBs display an additional absorption at $\lambda\sim5810$\,\AA. The wavelength is consistent with that of a blend of two fainter DIBs \citep[$\lambda\lambda$5809.12,5809.96][]{Jenniskens94}. The detection of this additional absorption is encouraging. However, its magnitude is comparable to the standard deviation of the residuals and it will not be discussed further.

\subsection{Quantifying bias and uncertainties \label{secbias}}

We simulated the bias and uncertainties of the whole global procedure by means of mock spectra realistically matching the \texttt{STARLIGHT} input spectra.
The greatest source of uncertainty of the DIB measurements is a possible contamination by a stellar spectral feature at $\lambda$5782, with contributions from \ion{Fe}{i}, \ion{Cr}{i}, \ion{Cu}{i}, and \ion{Mg}{i} \citep{Worthey94}. The feature varies with both age and metallicity of the underlying stellar population. We measured the equivalent width of this feature in our data, obtaining typical values of between $\sim$650 and $\sim$800~m\AA.
Taking this into account, we used a linear combination of two spectra from the base library \citep{bru03} as starting point of our simulations. They have an equivalent width of $\sim$700~m\AA\, for this feature, and very different ages ($\sim$15~Myr and $\sim$2.5~Gyr), both with solar metallicity. We note that the specific selection of the stellar population is not critical as long as the equivalent width of the stellar absorption feature is comparable to that measured in our spectra.
Then we added two Gaussian functions at the position of the DIBs at $\lambda$5780 and $\lambda$5797, with varying equivalent widths in $\lambda$5780 from 0 to 500 m\AA.
The ratio between the two features varies with the environments \citep[see e.g.][]{Vos11,Cox17} with the  DIB at $\lambda$5797 being typically three to five times fainter.
We assumed a factor of three between these two DIBs in our mock spectra.
To estimate the uncertainties, the specific value of this ratio is not critical so long as the whole range of expected equivalent widths are covered.
Next, the spectra were reddened using an attenuation law \citep{car89} and assuming a similar relation between extinction and the $\lambda$5780 DIB as in our Galaxy.
Following, we added random offsets in wavelength of up to $\pm1$~\AA, to take into account that the velocity of our spectra is a priori unknown by \texttt{STARLIGHT}.
Finally, we interpolated the simulated spectra to the sampling of the MUSE spectra, added some random noise to simulate the S/N of the \texttt{STARLIGHT} input spectra and scaled them so that the continuum around the DIBs is normalised to one.
The total number of mock spectra was 500. They were processed as described in Sect.~\ref{secmodelling} and Sect.~\ref{secfit}, exactly as our MUSE spectra.

The comparison between the input and measured equivalent widths for both DIBs in the mock spectra is shown in Fig.~\ref{figcomapmockspec}.
The figure shows that we are able to adequately recover the equivalent width for the DIB at $\lambda$5780 with a standard deviation for the residuals of $\sim40$~m\AA.
Still, at low equivalent width, our whole procedure systematically overestimates the equivalent width. In that sense, measured values $\lesssim$100 m\AA\, should be seen as upper limits.
Results are slightly different for the DIB at $\lambda$5797 (standard deviation of $\sim25$~m\AA). This feature is intrinsically fainter and thus more sensitive to our procedure. The experiment shows that our measuring procedure systematically overestimates the equivalent width for this DIB. The comparison of input and output equivalent widths provides us with a correction factor (the one-degree polynomial displayed as an orange continuous line in Fig. \ref{figcomapmockspec}) for our measurements.
We automatically rejected any measured equivalent width smaller than the standard deviation plus the zero-point for the one-degree polynomial fit between the input and output equivalent widths.
This limit is shown by the green dotted line in Fig.~\ref{figcomapmockspec}.
Given the faintness of the spectral features, all the remaining fits were inspected afterwards to avoid the inclusion of spurious DIB detection in the analysis.
The remaining measurements are discussed in following sections.
Maps of the DIB distributions are a direct presentation of our measurements without any correction, while the comparisons of the DIB strengths with other properties (i.e. reddening, measurements from literature) include a correction based on the one-degree polynomials derived above (i.e. orange line in Fig.~ \ref{figcomapmockspec}).

\begin{figure}[!th]
\centering
\includegraphics[angle=0,width=0.48\textwidth, clip=, viewport=35 75 500 490,]{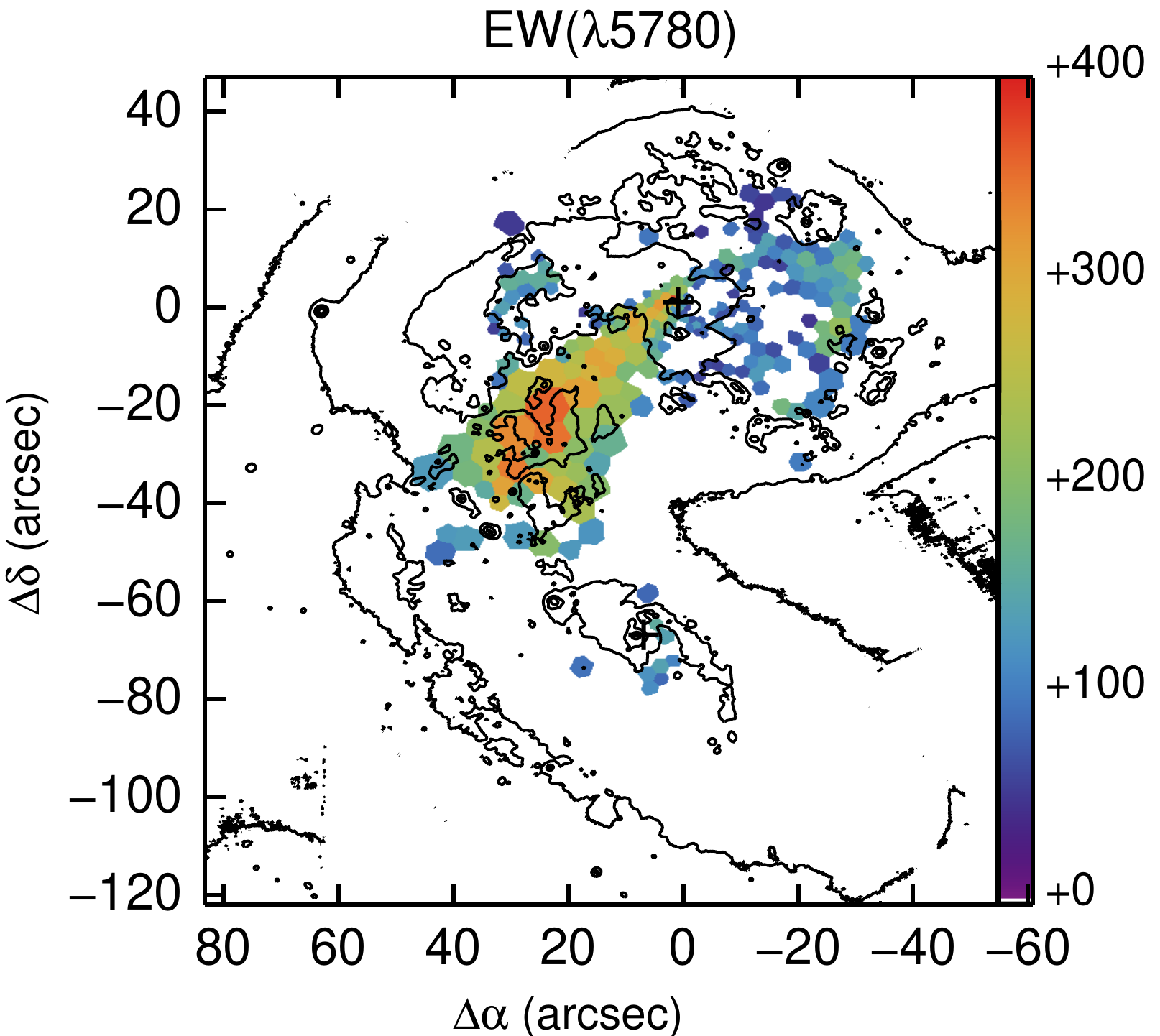}
\includegraphics[angle=0,width=0.48\textwidth, clip=, viewport=35 75 500 490,]{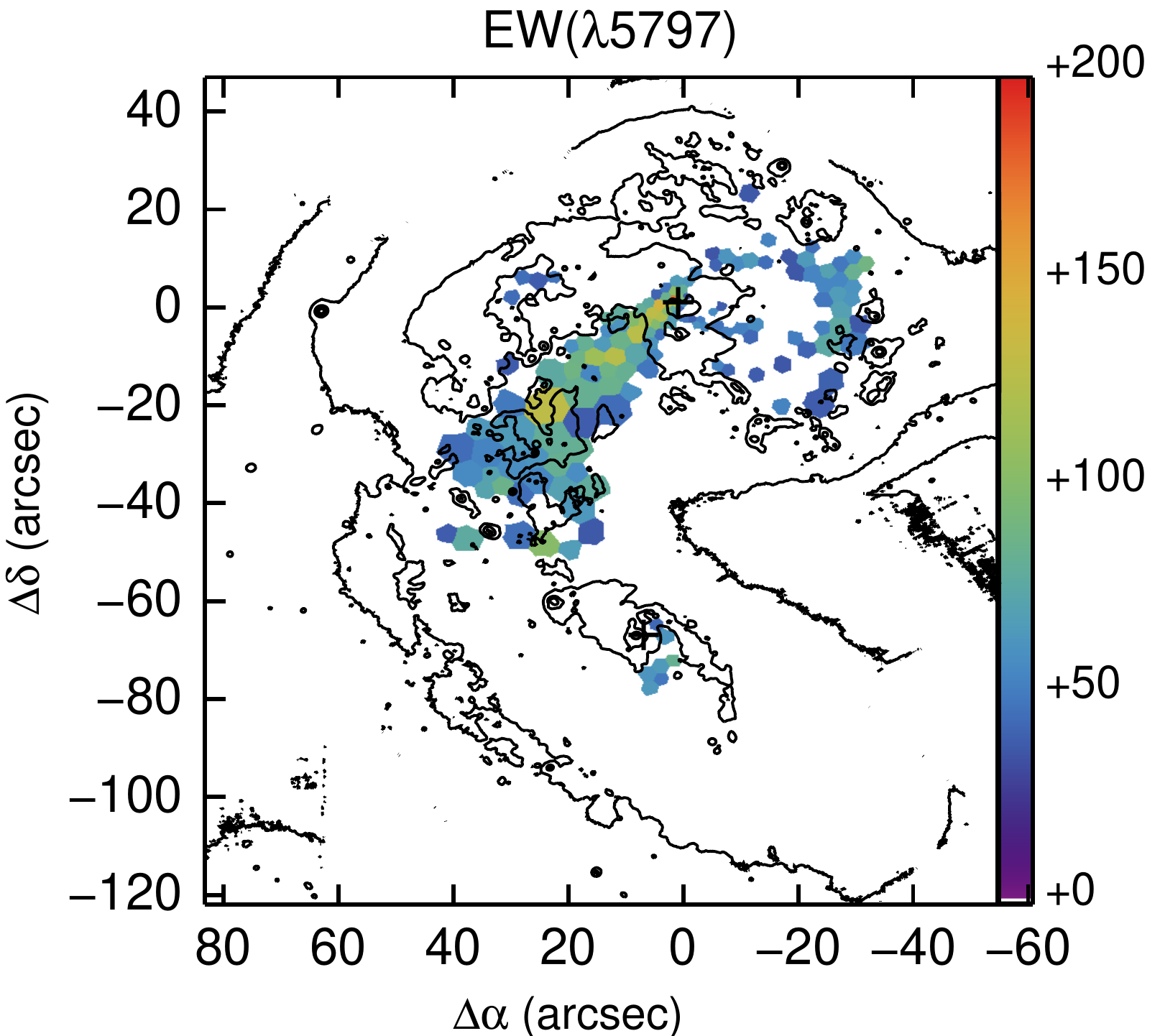}
\caption[]{Maps of the derived equivalent width in m\AA\, for DIBs at $\lambda$5780 (\emph{top}) and $\lambda$5797 (\emph{bottom}).
 The reconstructed white-light image is overplotted as reference with  contours in logarithmic stretching of 0.5 dex steps.
  }
\label{figdibmap}
\end{figure}

\section{Results and discussion}\label{sec:disc}

\subsection{Mapping the DIBs in the system}

The maps showing the detections of DIBs at $\lambda$5780 and $\lambda$5797 are presented in Fig.~\ref{figdibmap}. 
They constitute the main outcome of the present contribution.
Most of the $\lambda$5780 DIB absorption is found in a triangular area of $\sim0.6\sq\arcmin$ in the overlapping region of the system with the maximum ($\sim$340~m\AA) at its centre. Additionally, this DIB is also detected in several smaller locations in both galaxies. 
The structure of this distribution shows an
uncanny resemblance
to that of the obscured areas in the system, 
when compared with multi-band high spatial resolution with the HST \citep[see Fig.~2 of ][]{Whitmore10}, tracing not only the overlap area, but also an important part of the northern spiral arm. This suggests a relationship with the extinction, as has been found in the Milky Way and other nearby galaxies \citep[see e.g.][and references therein]{Welty14}
and will be explored in detail in Sects.~\ref{secdibandebv} and \ref{secelsewhere}.

As reviewed in the introduction, studies mostly in the Milky Way, but also in nearby galaxies, show that DIBs may or may not be related with other constituents of the ISM. In that sense, we can profit from the wealth of ancillary information about the \object{Antennae Galaxy} to explore if similar connections in this system exist. This is presented in the following sections.

\subsection{Neutral atomic hydrogen and DIBs}

\begin{figure}[th]
\centering
\includegraphics[angle=0,width=0.48\textwidth, clip=, viewport=35 75 500 490,]{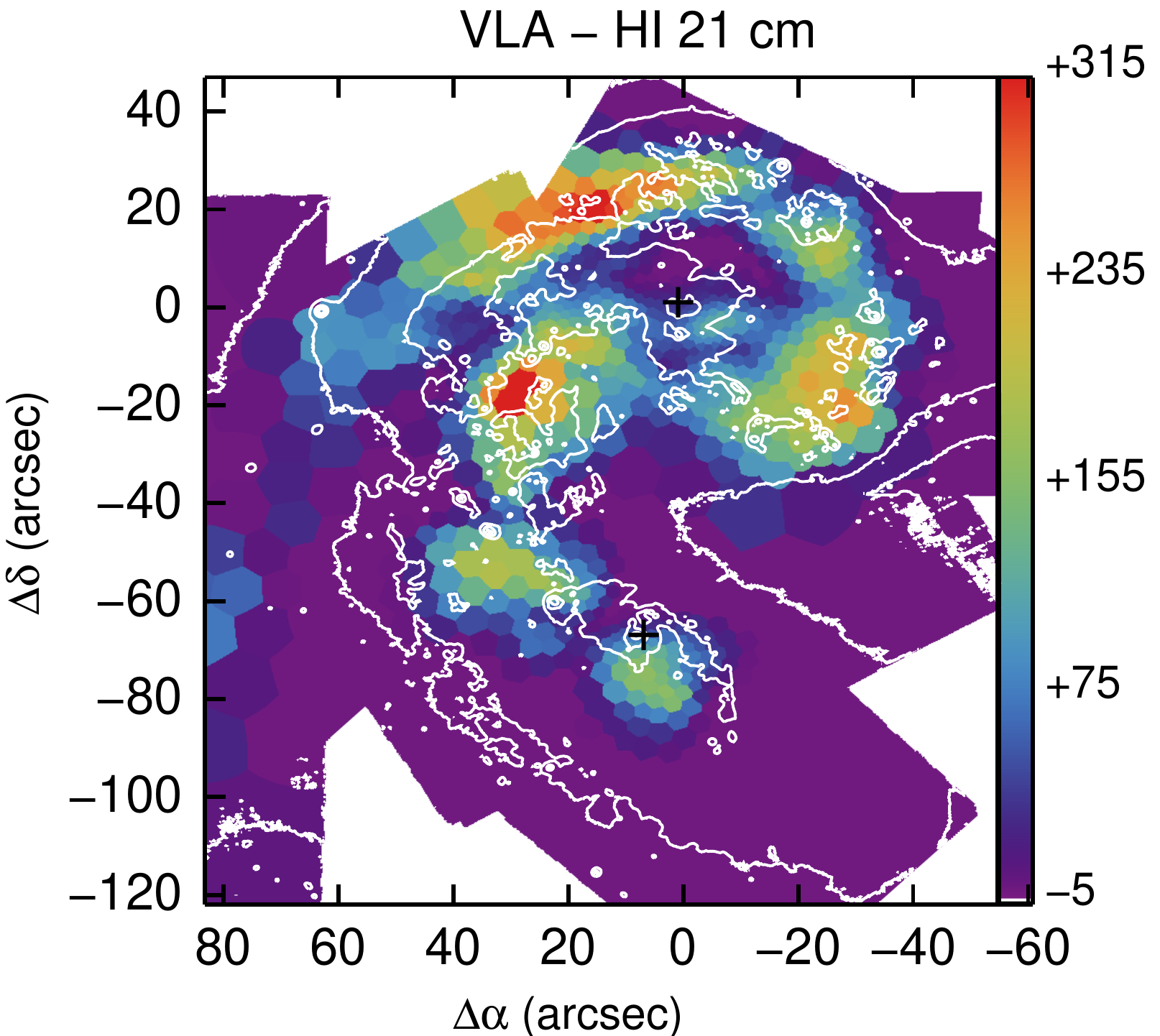}
\caption[]{
VLA \ion{H}{i} 21 cm map \citep{Hibbard01} once resampled and binned. Units are arbitrary and the colour scale is in linear stretch.
The reconstructed white-light image is overplotted as reference with  contours in logarithmic stretching of 0.5 dex steps. 
}
\label{figHI}
\end{figure}

The $\lambda$5780 DIB correlates well with neutral hydrogen in our Galaxy \citep[e.g.][]{Friedman11} while for a given amout of neutral hydrogen, DIBs seems underabundant in the Magellanic Clouds \citep{Welty06}.
Figure \ref{figHI} shows the \ion{H}{i} 21 cm line map with the VLA downloaded from the NED\footnote{\texttt{https://ned.ipac.caltech.edu/}}. The original map has a spatial resolution of 11\farcs40 \citep{Hibbard01}. Here the map has been resampled and binned matching our tessellation, to ease the comparison with the DIB maps. Looking at this map together with those presented in Fig.~\ref{figdibmap} (and also Fig.~\ref{apuntado}), we see that both the atomic hydrogen and DIB distributions do not necessarily match that of the stars, as seen in the optical. Interestingly, the $\lambda$5780 and $\lambda$5797 DIBs are detected in regions with an excess of neutral hydrogen (part of the overlap area, northern spiral arm), supporting the connection between the carriers of these two DIBs and \ion{H}{i}.
Still, the atomic hydrogen extends well beyond the region with detection of DIB absorption, as if the characteristic scale lengths for these were smaller. That is: at a given location, existence of \ion{H}{i}  seems a necessary condition for the existence of DIB carriers, but not enough to ensure it alone.
Similar results have been found in our Galaxy when comparing radial profiles of these species \citep{Puspitarini15}.
An exception is a narrow tongue at $\sim[-10^{\prime\prime},-2^{\prime\prime}]$ joining the northern nucleus with the overlap area. While absorption in both DIBs have been detected, the VLA \ion{H}{i} 21 cm does not show any particularly strong emission there. The spatial resolution of the VLA image, larger than the width of the tongue, may explain this effect. Our DIB detections and the VLA map would still be compatible with the need of \ion{H}{i} for having DIB absorption if at this location there were  a localised concentration of \ion{H}{i} with typical size larger than the width of the tongue but smaller than the VLA resolution.

\begin{figure}[th]
\centering
\includegraphics[angle=0,width=0.48\textwidth, clip=, viewport=35 75 500 490,]{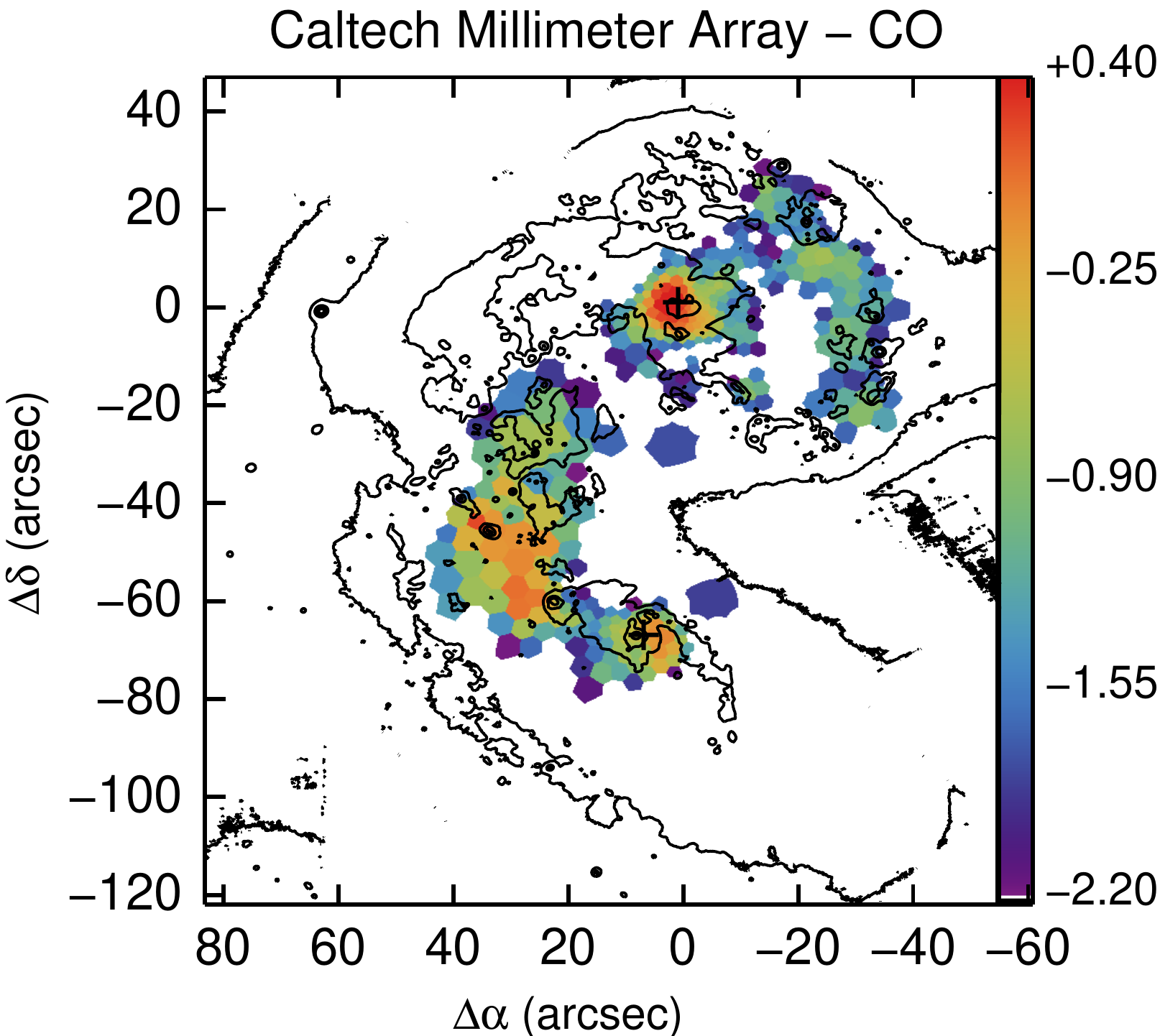}
\caption[]{
Caltech Millimeter Array CO map \citep{Wilson00} once resampled and binned. Units are arbitrary and the colour scale is in logarithmic stretch.
The reconstructed white-light image is overplotted as reference with  contours in logarithmic stretching of 0.5 dex steps.
}
\label{figCO}
\end{figure}

\subsection{Molecular gas and DIBs\label{secmolecgas}}

ALMA observations of the Antennae Galaxy in Carbon monoxide (CO) are available \citep{Whitmore14}. However, they do not cover the whole area of interest, and in particular, the region with high DIB absorption. Therefore, we preferred to use the Caltech Millimeter Array CO map presented by \citet{Wilson00}.
The observations were taken with a synthesised beam of 3\farcs15$\times$4\farcs91. 
A resampled and binned version of this CO map is presented in Fig.~\ref{figCO}.

The cold molecular gas distribution in the system, as traced by the CO emission, peaks at the northern nucleus, is relatively strong at the southern one and presents three important local maxima at roughly [$30^{\prime\prime},-60^{\prime\prime}$], [$30^{\prime\prime},-50^{\prime\prime}$] and [$35^{\prime\prime},-45^{\prime\prime}$] where the two disks overlap
\citep{Wilson00,Zhu03,Smith07,ZaragozaCardiel14}. 
This does coincide with the location where
the most intense star formation in the system occurs \citep{Mirabel98,Vigroux96}. The warm molecular gas, as traced by the H$_2$ S(2) and S(3) rotational lines, displays a similar spatial structure \citep{Brandl09}.

Molecular gas and DIBs display quite different distributions. 
With the exception of some marginal DIB detection close to the northern nucleus, we find no DIB absorption up to our threshold in places with strong molecular gas emission.
This is in line with results for the Milky Way that show that DIB carriers tend to disappear in dense cores of clouds \citep{Lan15} - the so-called  skin effect\citep{Snow74} - and that there is a lack of correlation between DIB strength and molecular gas \citep{Friedman11,Welty14b}.
Still, the maximum of the DIB absorption is found in a region with mild emission in CO, and both DIBs and CO are detected in the northern spiral arm.

The lack of DIB detection at the location with strong CO emission coupled with the detection in the places with mild CO emission in \object{Arp\,244} could be explained in the context of the \emph{skin effect}, as in the Milky Way. This is interpreted as being that DIB formation takes place most efficiently in the outer regions of the interstellar cloud.
Additionally, extinction could play a role, as will be discussed in Sect. \ref{secdibandebv}.

\begin{figure}[th]
\centering
\includegraphics[angle=0,width=0.48\textwidth, clip=, viewport=35 75 500 490,]{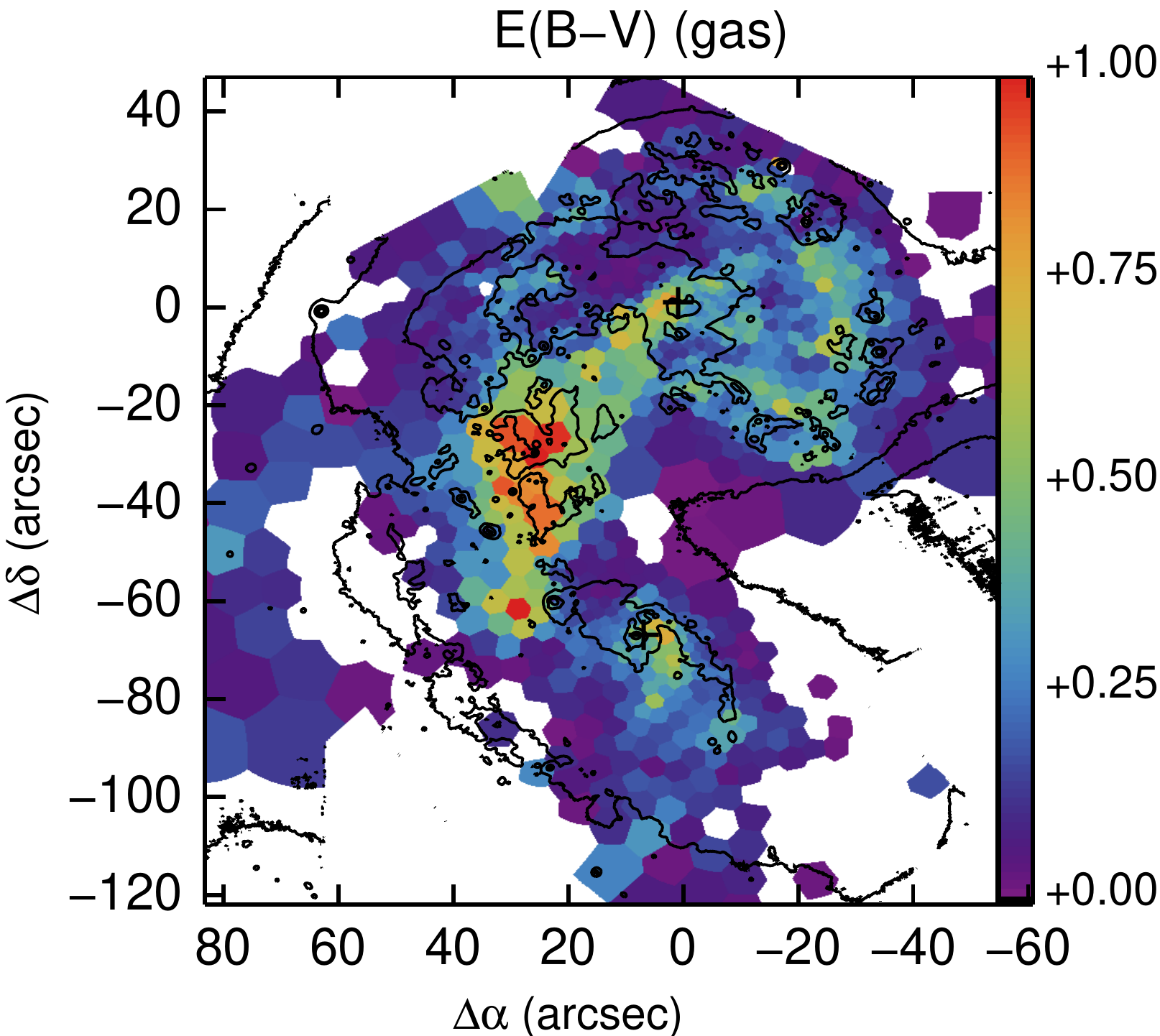}
\includegraphics[angle=0,width=0.48\textwidth, clip=, viewport=35 75 500 490,]{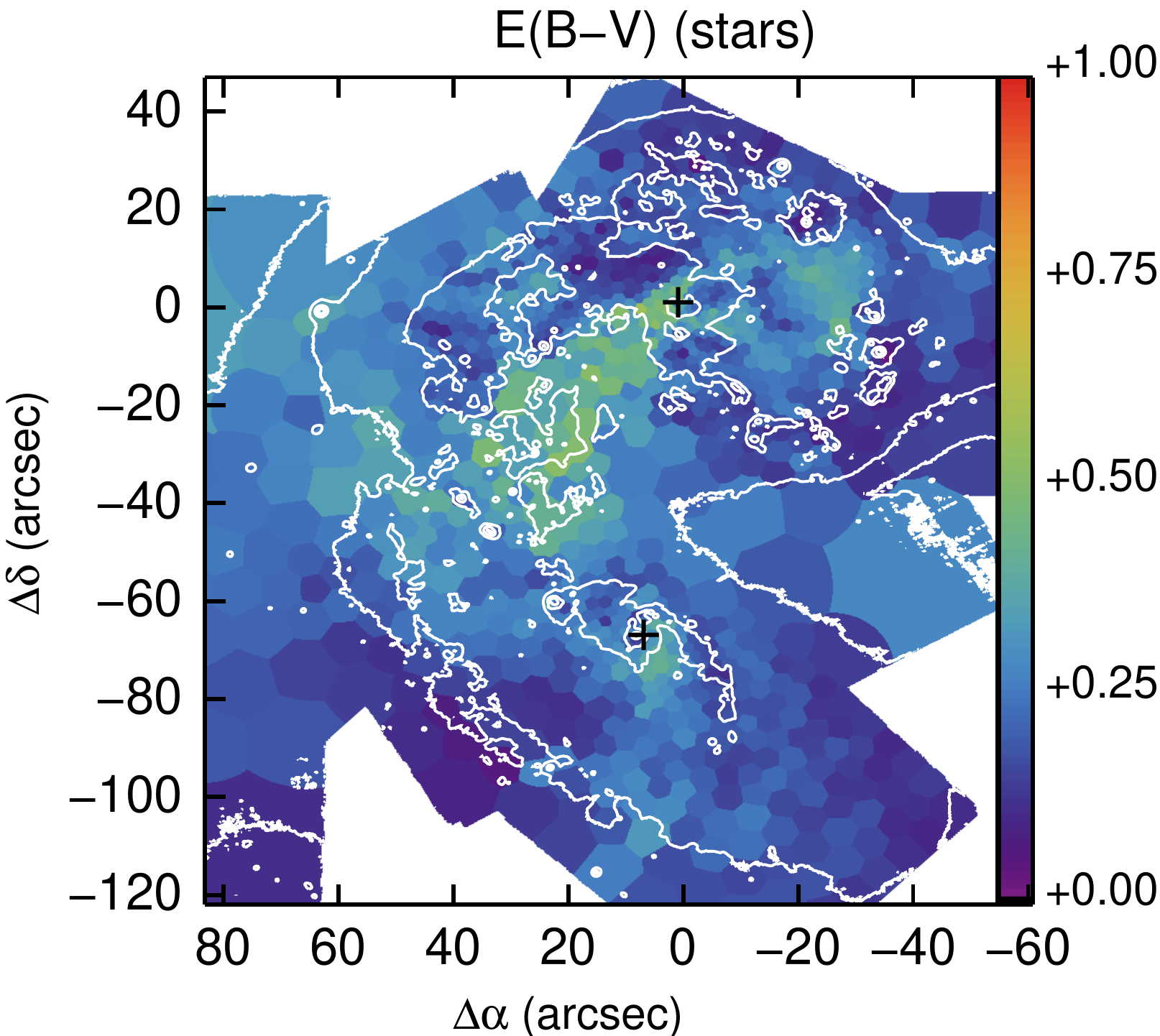}
\caption[]{Derived reddening maps for the ionised gas (\emph{top}) and the stars (\emph{bottom}).
 The reconstructed white-light image is overplotted as reference with  contours in logarithmic stretching of 0.5 dex steps. 
}
\label{figebvmap}
\end{figure}

\subsection{Attenuation and DIBs\label{secdibandebv}}

It is well established that the DIB strength presents a good correlation with the interstellar extinction in the Milky Way \citep[see e.g.][and references therein]{Capitanio17,Lan15} and other galaxies \citep[e.g.][]{Cordiner11}.
Here, we obtained a total number of independent measurements of DIB strength within the system of $>200$ for the $\lambda$5780 DIB and $>100$ for the  $\lambda$5797 DIB. 
This is comparable to the number of lines of sight typically used in many Galactic studies.
Thus, we are in an optimal position to explore whether a similar relationship exists also in an environment experiencing a star formation as extreme as that found in starburst galaxies.

\begin{figure}[th]
\centering
\includegraphics[angle=0,width=0.24\textwidth, clip=, viewport=0 0 430 300,]{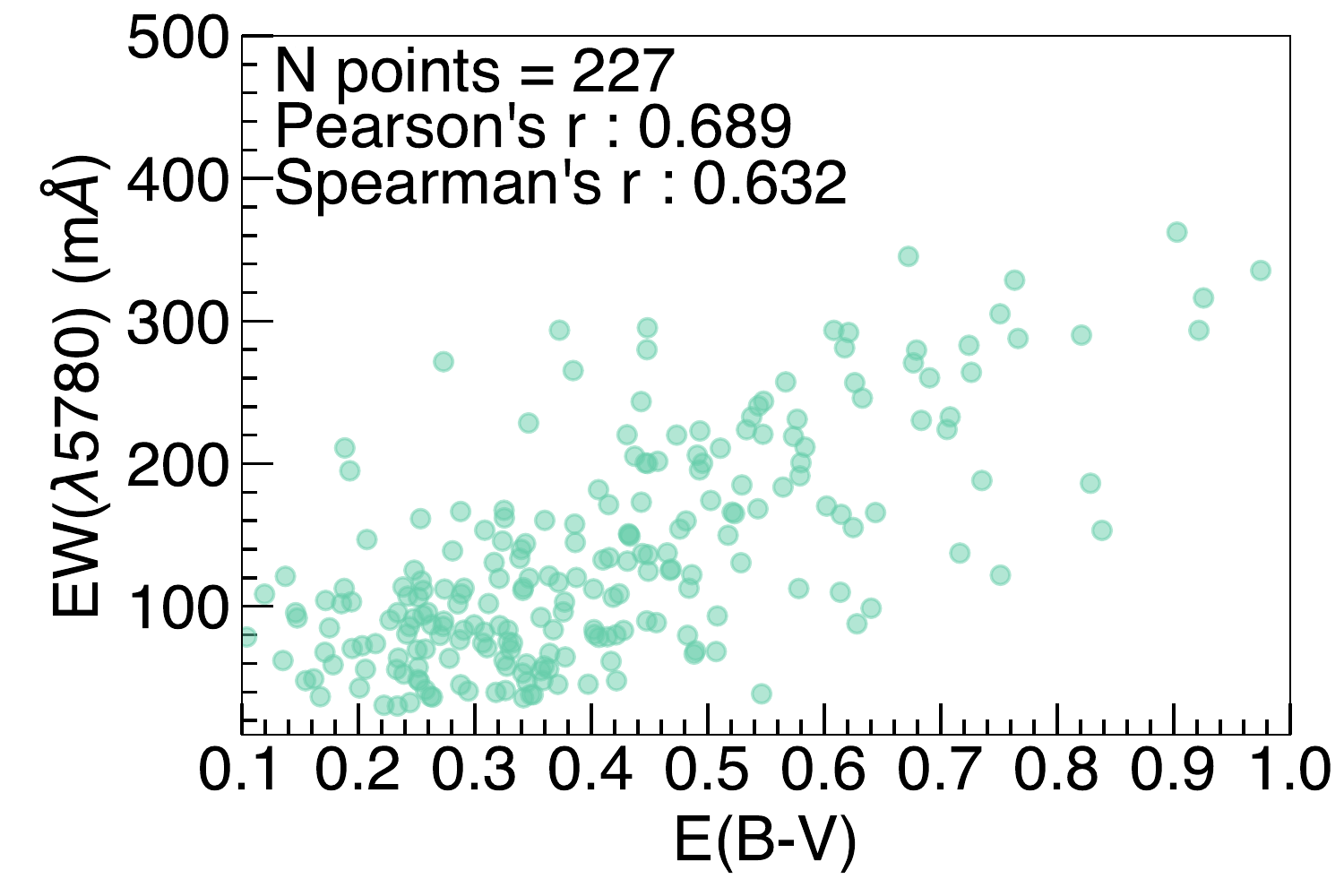}
\includegraphics[angle=0,width=0.24\textwidth, clip=, viewport=0 0 430 300,]{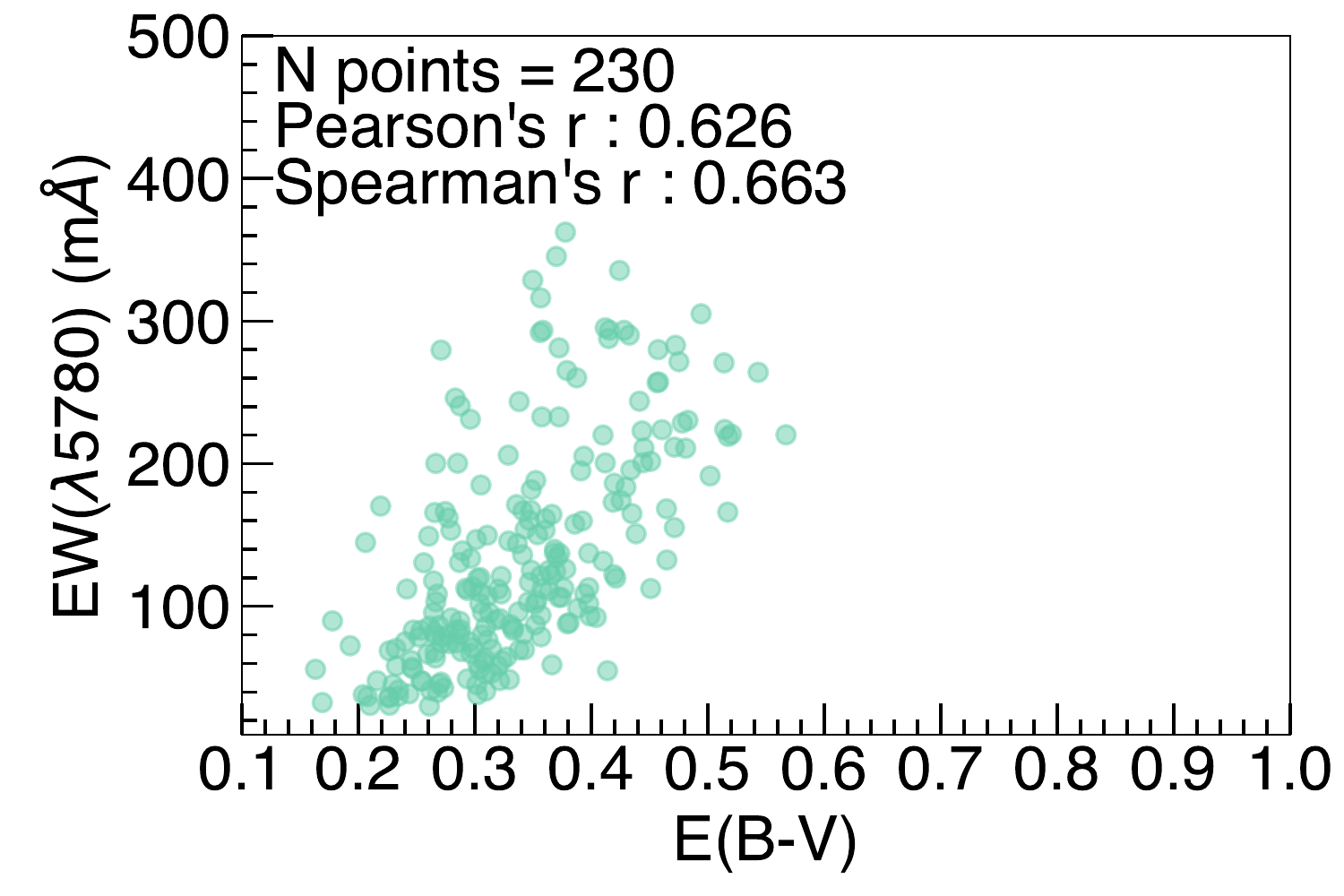}
\includegraphics[angle=0,width=0.24\textwidth, clip=, viewport=0 0 430 300,]{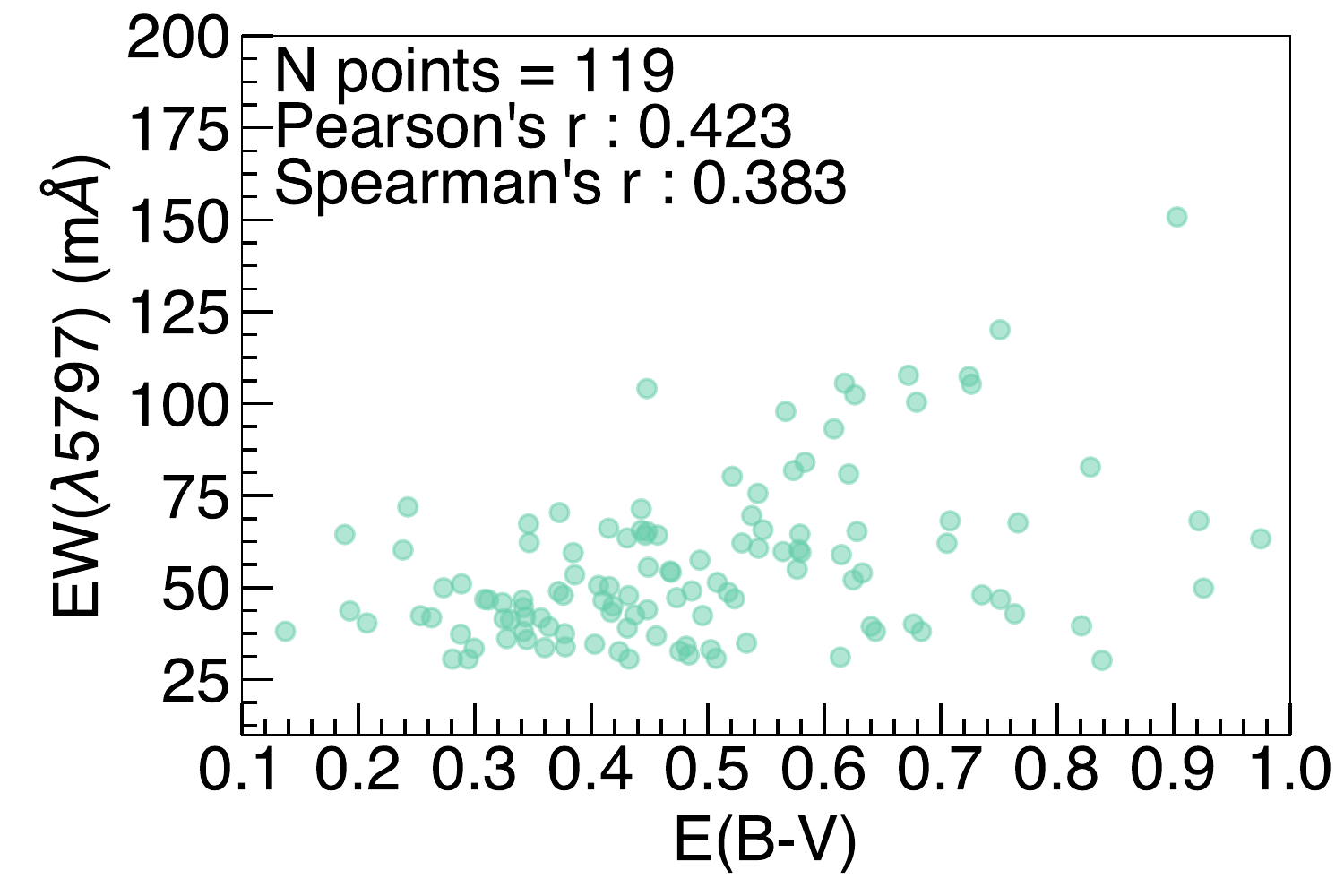}
\includegraphics[angle=0,width=0.24\textwidth, clip=, viewport=0 0 430 300,]{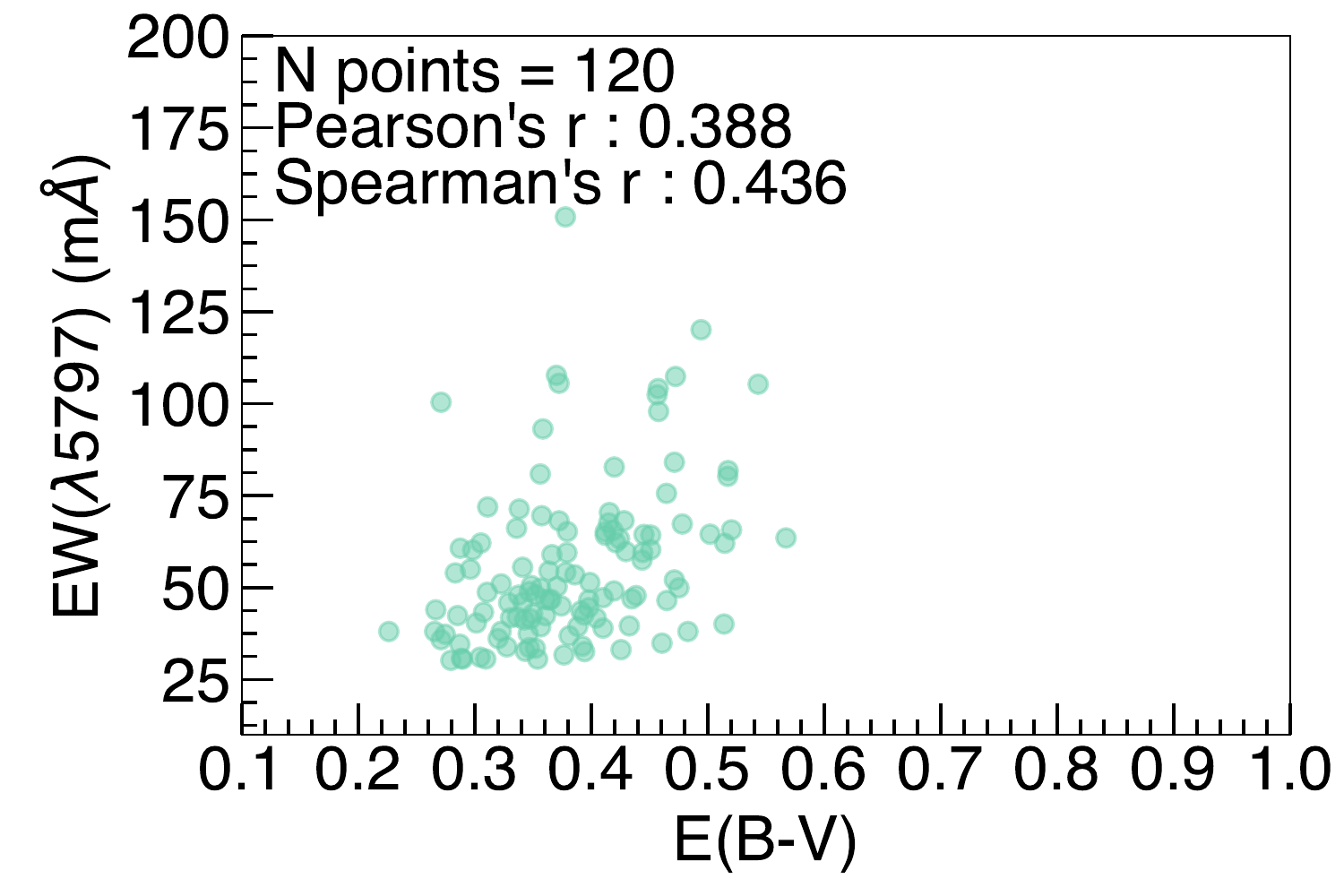}
\caption[]{Comparison between the strength of the DIBs at  $\lambda$5780 (\emph{top}) and at $\lambda$5797 (\emph{bottom}) and the reddening of the ionised gas (\emph{left}) and the stars (\emph{right}).
Uncertainties for the gas reddening range typically between 0.06 and 0.14~dex. Uncertainties for the $\lambda$5780 and  $\lambda$5797 equivalent widths were $\sim$50 and  $\sim$40~m\AA, respectively (see Sec. \ref{secbias}).
}
\label{figebvvsdibgal}
\end{figure}

We present two reddening maps for the Antennae Galaxy in Figure \ref{figebvmap}.
They are not corrected for Galactic extinction.
In the direction of the Antennae, this is $A_V = 0.127$  \citep{Schlafly11}.
Assuming a selective reddening of $R_V = 3.1$ \citep{Rieke85}, this implies an \ebv=0.041. 
In galaxies where one cannot resolve individual stars, the most similar accessible magnitude to the extinction is luminosity-weighted attenuation, that can present some differences due to transfer effects and scattering into the line of sight. Still, these differences in the attenuation and extinction curves are mostly found in the ultraviolet being essentially the same in the optical \citep{Gordon03}, and more specifically in the wavelength range used in this work. Thus we will assume here that extinction laws derived for our Galaxy can be used to discuss attenuation in the optical within the system.
The map in the upper panel was derived using the extinction curve presented by \citet{Fluks94} and assuming an intrinsic Balmer emission-line ratio \ha/\hb = 2.86 for a case B approximation and T$_e$ = 10\,000~K \citep{Osterbrock06}. As such, it represents the attenuation towards the ionised gas (and young stellar population). The one in the lower panel was provided by \texttt{STARLIGHT}. It represents the attenuation of the overall stellar population at optical wavelengths. In both cases, attenuation is high in the overlap region, being higher when using the \ha \ to \hb\, ratio. This has already been seen in other starbursts and can be explained if the dust has a larger covering factor for ionised gas than for the stars \citep[e.g.][]{Calzetti97,MonrealIbero10a}.

When comparing both maps with those in Fig.~\ref{figdibmap} a clear result arises: the overall structure is comparable and DIBs are detected there where attenuation is high.
The highest attenuation for both, ionised gas and overall stellar population, does coincide with the location with largest DIB absorption (at $\sim[25^{\prime\prime},-25^{\prime\prime}]$ in our relative coordinates system).
Still, in the area with highest concentration of molecular gas (and no DIB detection), the overall stellar population as seen in the optical is subject of relatively low attenuation while that derived for the ionised gas is relatively high.
This can be understood in a context where the optical stellar continuum is blind to the very heart of the overlap region where most of the massive star formation takes place. $E(B-V)$ values provided by \texttt{STARLIGHT} and based on the stellar continuum trace only the most superficial layers in this area. 
Then, in these layers the concentration of DIBs carriers should be low, too.
Together with the previously mentioned skin effect, this may explain the lack of DIB absorption in this area. 
We will come back to this point when discussing the relation between DIB absorption and emission in the mid-infrared band (see Sect.~\ref{secpah}).

The relationship between DIBs and attenuation is further explored in Fig.~\ref{figebvvsdibgal}. The upper row shows the results for the $\lambda$5780 DIB. In the locations where the DIB was detected, it correlates equally well with both the stellar and ionised gas attenuation, in the sense that both the Pearson and the Spearman correlation coefficients are comparable. The correlation is clear and significant according to a Student's t-test, although the coefficients are not as high as those determined for our Galaxy \citep[typically using much higher spatial and spectral resolution, e.g.][]{Friedman11,Puspitarini13}. Actually, the difference in spectral resolution can at least partially explain the lower strength of the correlation for the Antennae Galaxy. The $\lambda$5780 is blended with another broader DIB at $\lambda$5779 \citep{Jenniskens94}. Here, since the spectral resolution is relatively low, we did not make any attempt to separate both features, and thus variation in their relative contributions to the measured equivalent width can have an impact on the robustness of the $E(B-V)$ - EW($\lambda$5780) correlation.

A relationship between the $\lambda$5797 DIB and attenuation is not that clear from the present data. Correlation coefficients are low, and according to them only about $20-25$\% of the variation in the EW($\lambda$5797) is somehow related to the amount of reddening. This is in clear contrast with results for the Milky Way, for which this DIB is an even better proxy for the extinction than the one at $\lambda$5780 \citep{Ensor17}.
However, these low correlation coefficients are largely weighted by a series of measurements along several lines of sight with EW($\lambda$5797) of only about two times the estimated uncertainty.
Most of the data points in fact lie between 40 and 100~m\AA. Thus we are mostly sampling a range in equivalent widths comparable to the estimated uncertainties, implying a meagre contrast to detect a solid correlation.
Moreover, this DIB is much narrower than that at $\lambda$5780, and is also blended with a broader DIB at $\lambda$5795.
In that sense it is more sensitive to the spectral resolution of the instrument than the $\lambda$5780 DIB.

On the positive side, when looking at high EW($\lambda$5797)s, there is a clear tendency in the sense that lines of sight with larger equivalent width also have higher gas extinction. This is encouraging and calls for a similar experiment (in this or another galaxy) with an IFS similar to MUSE but with higher spectral resolution. This is exactly what the spectrographs foreseen for the future ESO ELT, HARMONI, and MOSAIC, will be able to provide. In the meantime, instruments such as MEGARA on GTC could also be used to obtain data with reduced uncertainties and covering a larger range of equivalent widths.

To find out whether these measurements support a putative correlation between this DIB and extinction, we can compare our results with those found for other galaxies, that sample a larger range of equivalent widths. This is the subject of the discussion in Sect.~\ref{secelsewhere}.

\subsection{Attenuation and DIBs here and elsewhere \label{secelsewhere}}

In this section, we discuss how the relationships between reddening and strength of the DIBs found in the previous section compares with those found in other galaxies.

\begin{figure}[!th]
\centering
\includegraphics[angle=0,width=0.49\textwidth, clip=, viewport=5 0 570 280]{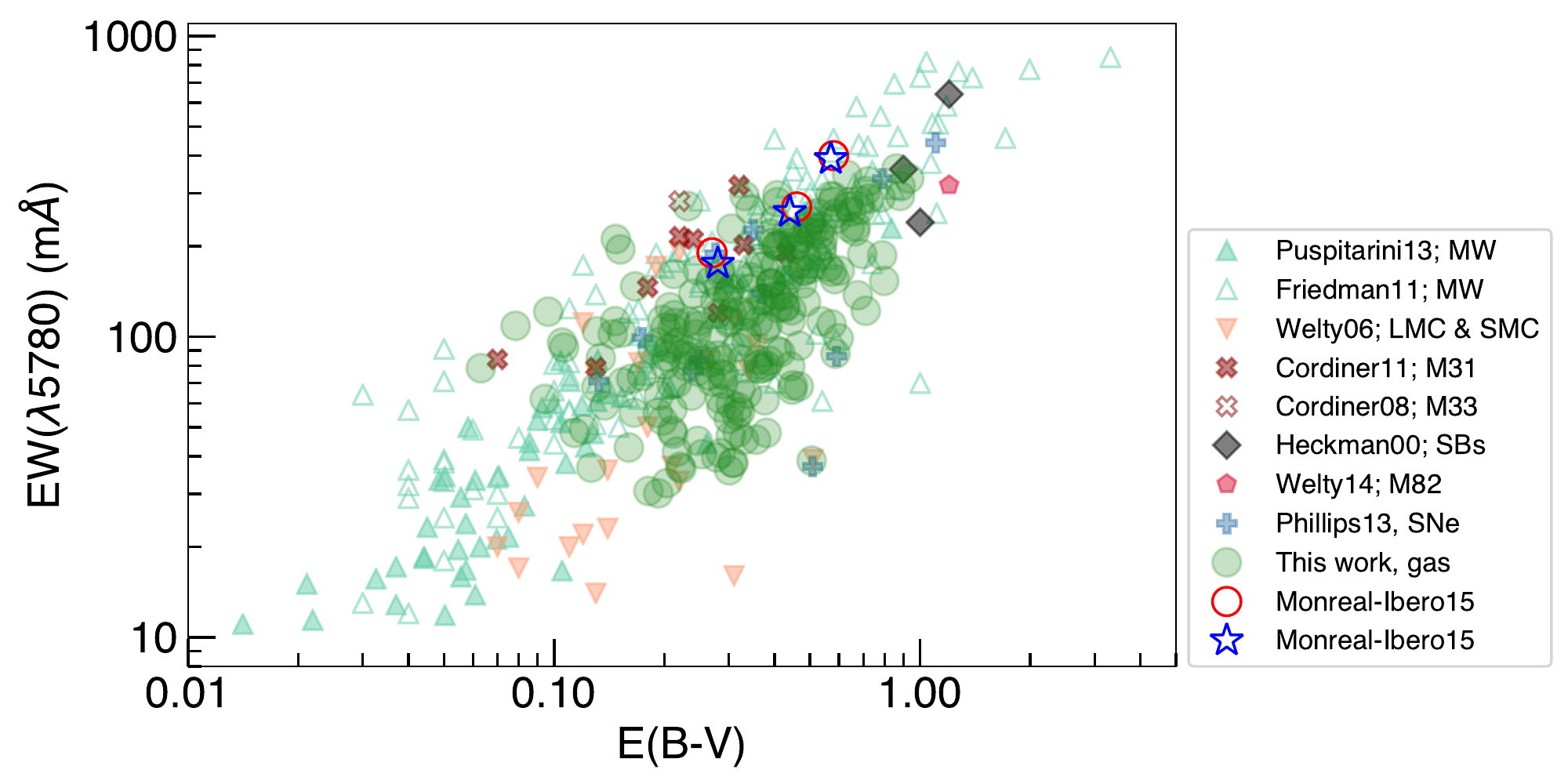}
\caption[]{Relation between the strength of the $\lambda$5780 DIB and the reddening for the \object{Antennae Galaxy} (green circles for the gas extinction), and other galaxies.
For the \object{Antennae Galaxy} data, a Galactic reddening $E(B-V)_\mathrm{gal}=0.041$ has been subtracted.
Regarding the data for other galaxies, we included two sets of observations for Galactic targets (blue triangles; filled: \citet{Puspitarini13}, open: \citet{Friedman11}), several sets of data for galaxies in the Local Group (inverted salmon triangles: Magellanic Clouds \citep{Welty06}; filled burgundy "x": M\,31 \citep{Cordiner11}, open burgundy "x": M\,33 \citep{Cordiner08a}), a set of measurements through dusty starbursts (black diamonds, \citet{Heckman00}), some measurements in the lines of sight of supernovae \citep[filled blue crosses,][]{Phillips13}, including one in \object{M\,82} \citep[coral pink pentagon,][]{Welty14} and our previous  results for \object{AM\,1353-272}~B \citep[red circles and blue stars,][]{MonrealIbero15}.
}
\label{figothergal5780}
\end{figure}

This is summarised in Fig.~\ref{figothergal5780} for the $\lambda$5780 DIB. The figure is similar to Fig.~4 by \citet{MonrealIbero15}, but with the new data for the \object{Antennae Galaxy} overplotted.
It illustrates beautifully the enormous potential of IFSs for extragalactic DIB research: passing from a few tens to a few hundreds of extragalactic measurements of the $\lambda$5780 DIB was possible by adding only one set of observations. 
As with \object{AM\,1353-272}~B, most of our measurements for \object{Arp 244} fall very nicely in the relation defined by our Galaxy, \object{M\,31}, and \object{M\,33} all of them spiral galaxies with solar or near-solar metallicities \citep[e.g.][]{ToribioSanCipriano16,Lardo15,Sanders12}\footnote{Here we assume $12+\log(O/H)_\odot=8.69$ \citep{Asplund09}.}.
and separated from the relation defined by the Magellanic Clouds\footnote{Metallicities for the LMC and SMC are $12+\log(O/H) = 8.35$ and 8.03 \citep{Russell92}.}. Thus, our results support the existence of a second parameter (e.g.\ metallicity, the characteristics of the radiation field) modulating the $E(B-V) -$ EW($\lambda$5780) relation. 
Similar observations to those presented here for additional dwarf(-ish) galaxies at sub-solar metallicities (for example) would be instrumental to further discuss the relative role of the possible second parameter. 

\begin{figure}[!th]
\centering
\includegraphics[angle=0,width=0.49\textwidth, clip=, viewport=5 0 570 280]{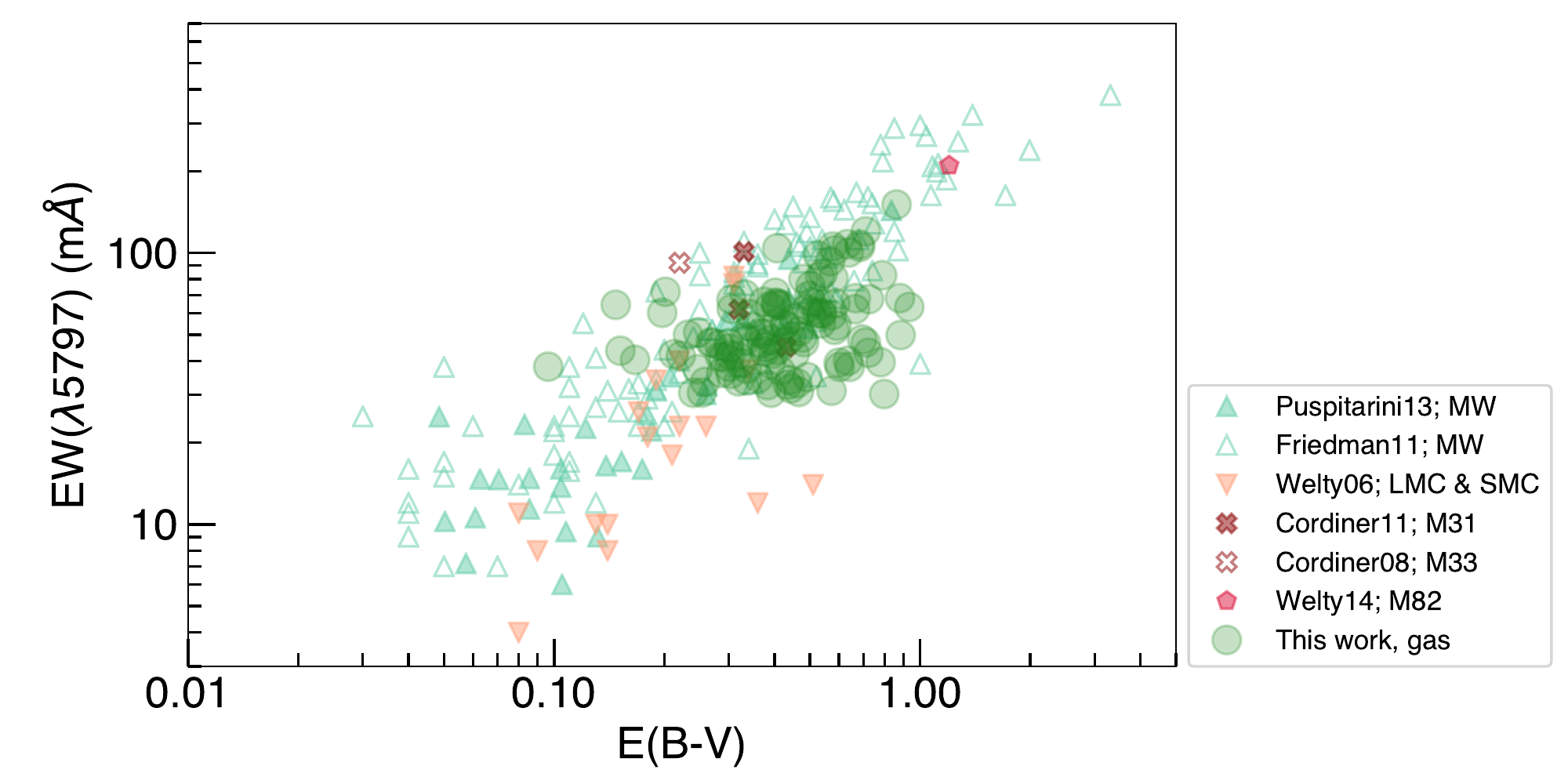}
\caption[]{Same as Fig.~\ref{figothergal5780} but for the $\lambda$5797 DIB. We note that results of  \citet{MonrealIbero15} and \citet{Phillips13} are not included, since these authors did not measure the $\lambda$5797 DIB.
}
\label{figothergal5797}
\end{figure}

Figure~\ref{figothergal5797} contains a similar relation to that in Fig.~\ref{figothergal5780} but for the $\lambda$5797 DIB.
Even if the data uncertainty is large, here we see that this DIB actually falls in the locus defined by measurements in other galaxies (including the \object{Milky Way}) and that when a larger range in equivalent widths is considered, the correlation between this DIB and reddening is unveiled.
Interestingly enough, the separation between data for the Magellanic Clouds and for galaxies with solar or solar-like metallicities is not as neat for this DIB as for the $\lambda$5780 DIB. This difference in behaviour is in line with the idea of different carriers for these two absorption bands
and in harmony with results that show, for example,\ the different reactions of these two DIBs to the radiation field \citep[e.g.][]{Vos11,Cami97}.

\begin{figure}[th]
\centering
\includegraphics[angle=0,width=0.48\textwidth, clip=, viewport=35 75 500 490,]{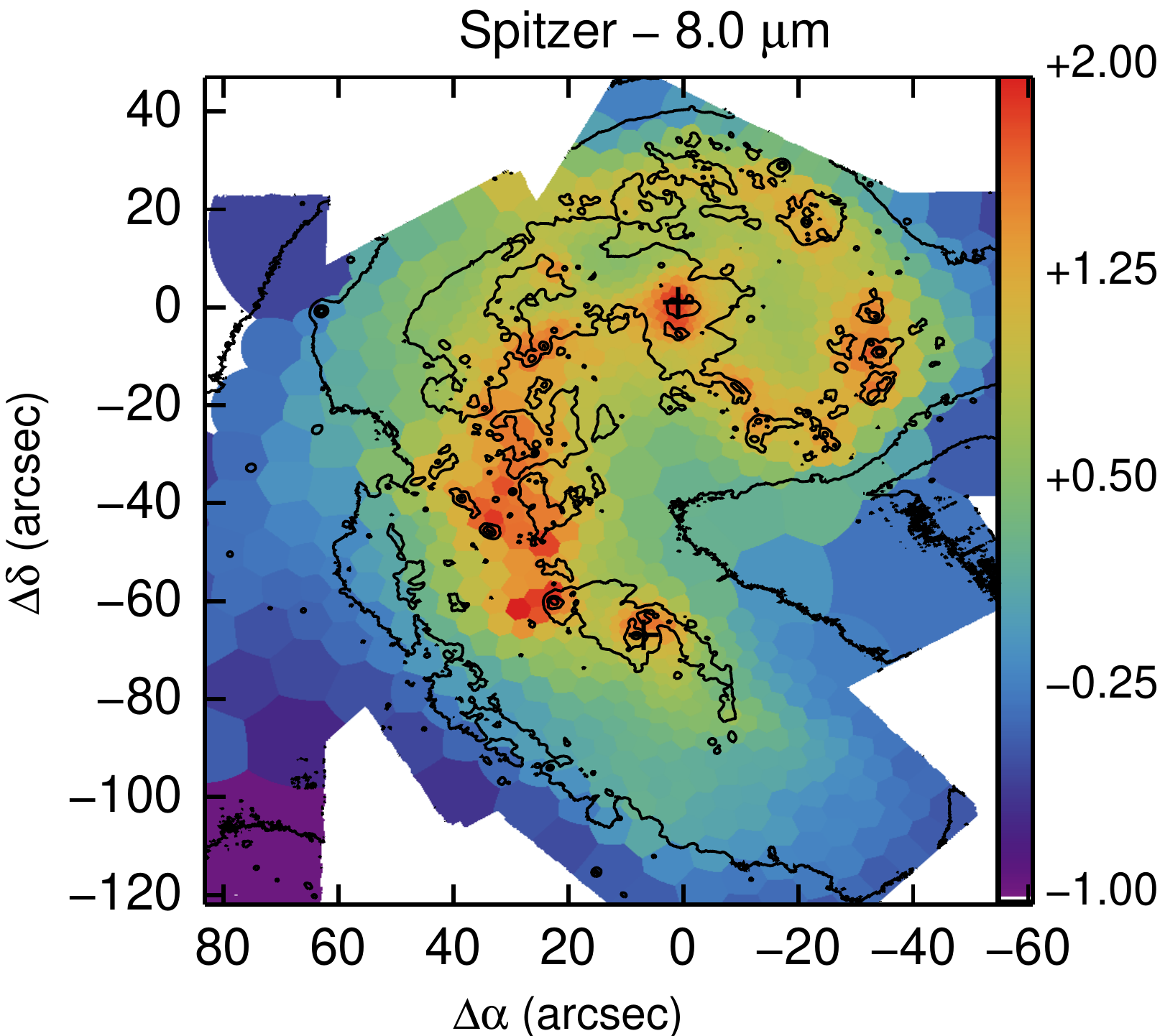}
\caption[]{
Spitzer - IRAC image in the 8 $\mu$m band \citep{Smith07} once resampled and binned. Units are arbitrary and the colour scale is in logarithmic  stretch.
 The reconstructed white-light image is overplotted as reference with  contours in logarithmic stretching of 0.5 dex steps.
  }
\label{figspitzermap}
\end{figure}

\subsection{Mid-IR emission and DIBs \label{secpah}}

Neutral and/or ionised PAHs are among the favourite candidates to be (some of) the DIB carriers \citep{Salama96}.
Several strong emission bands in the mid-IR at, for example,\ 3.3, 6.2, 7.7, and 8.6 $\mu$m, the so-called  unidentified infrared emission bands (UIBs), are also believed to be caused by PAHs and related species \citep[e.g.][]{Rigopoulou99,Draine07}.\footnote{UIBs are often directly referred in the literature as the PAH bands or PAH emission. Here, we reserve the acronym UIBs to refer to the spectroscopic features and the acronym PAHs to refer to the molecules themselves, to avoid ambiguities.}
Thus, in the context of gaining insight into the nature of the carriers, a comparison between our DIB maps and the emission at these wavelengths is an important test.

Figure~\ref{figspitzermap} shows the Spitzer - IRAC image in the 8.0~$\mu$m band downloaded from the NED \citep[][see also \citealt{Wang04}]{Smith07}.
The original spatial resolution was $\sim$1\farcs6.
As for the \ion{H}{i} 21 cm line and CO maps, it was re-sampled and binned to match our tessellation and ease the comparison with the DIB maps. This image basically reflects emission from the strong 7.7 and 8.6 $\mu$m bands, with minor (i.e.\ $<5\%$) stellar contribution (\citealt{Wang04}). 
A similar image for the 5.8~$\mu$m band -- that contains the strong 6.2~$\mu$m UIB -- is also available. We did not include it here since it displays a similar morphology to the one at 8.0~$\mu$m. 
Additionally, \citet{Brandl09}, by scanning with the \emph{Spitzer} - IRS, obtained spectral maps for the 8.6~$\mu$m and 11.3~$\mu$m UIBs (see their Fig.~4). All these maps display roughly the same structure: strong emission is seen in the two nuclei of the galaxies, the outer spiral arm of \object{NGC\,4038,} and the overlap region, with an irregular patchy structure.

We note that there is a marked difference between the structure of the UIBs (presumably caused by PAHs) and our DIB maps in Fig.~\ref{figdibmap}. While UIBs are strong over the whole overlap region, where atomic and especially molecular gas are strong, DIBs are detected only there where the \ion{H}{i} concentration is the highest and CO emission is moderate.
This result does not necessarily imply that PAHs or (PAH-related molecules) can be rejected as possible carriers of the $\lambda$5780 and $\lambda$5797 DIBs.
As we discussed in Sect.~\ref{secdibandebv}, in the area of high CO emission, optical and mid-infrared are actually sampling different depths along the line of sight. Should these two DIBs and the UIBs share related carriers, it is therefore reasonable to expect that their spectral signatures are inaccessible in the optical and can only be detected in the mid-infrared.
This would be a situation similar to that for the stars: most of the recent star formation is hidden in the optical \citep{Whitmore10} but revealed in the near and mid-infrared \citep{Brandl05,Wang04}.
In that context, it would be interesting to see whether infrared DIBs, less sensible to extinction effects are found in this area.

\section{Conclusions}\label{sec:concl}

This is the second of a series of papers aiming at a detailed study of the \object{Antennae Galaxy} with MUSE.
At the same time, it is the continuation of an experiment devoted to explore the potential of using IFSs in extragalactic DIB research.
In this contribution, we used high signal-to-noise ratio spectra of $\sim$1200 lines of sight to measure the absorption strength of the well-known $\lambda$5780 and $\lambda$5797 DIBs. We examined the correlations between them and the extinction in the system, and how their spatial distributions compare with that of other components of the ISM. Our main findings are:

\begin{enumerate}
\item To our knowledge, we have derived the first spatially resolved maps for the DIBs at $\lambda$5780 and $\lambda$5797 in galaxies outside the Local Group.
The detection area was largest for the strongest of the two DIBs (at $\lambda$5780) and covered a region of $\sim0.6\square^\prime$, corresponding to a linear area of $\sim25$~kpc$^2$.

\item Both DIBs correlate with atomic hydrogen in the sense that they were detected in the area with largest amount of this gas phase. However, typical length scales for these species are different, with atomic hydrogen extending well beyond the area presenting DIB absorption.
Thus, the existence of atomic hydrogen seems a necessary condition but not sufficient for the existence of DIBs.

\item DIBs and molecular gas present only mildly similar distributions. This is in accord with results for the Milky Way that indicates a lack of correlation between molecular gas and the $\lambda$5780 and $\lambda$5797 DIBs. DIBs are found in areas with moderately CO emission but they are absent in those with the highest CO emission. Both, the skin effect and the high extinction towards the areas with the highest CO emission can partially explain this.

\item The overall DIB and extinction distributions are comparable: the locations for the highest DIB absorption and highest extinction do coincide. However, large extinction for the ionised gas is seen also there where the concentration of molecular gas is maximal and no DIB absorption has been detected.

\item Within the system, the DIB $\lambda$5780 correlates well with both the gas and stellar extinction, similar to what happens with this DIB in the Milky Way.

\item The $\lambda$5797 DIB does not strongly correlate with the extinction. 
We attribute this mainly to a lack of contrast for the covered equivalent widths and estimated uncertainties. Still, data points with high EW($\lambda$5797) indicate that this connection may actually exist. 

\item When comparing our results with those for the EW($\lambda$5780) - $E(B-V)$ relation for other galaxies, the $\lambda$5780 DIB data falls in the sequence defined by other solar or solar-like spiral galaxies and far from the locus defined by the Magellanic Clouds.
The compilation of sets of data similar to these for an ensemble of galaxies with different metallicities would open the possibility of establishing relations of the kind of, for example, $E(B-V) = f(EW(DIB),Z)$. This would allow the use of DIBs as a proxy for the extinction in the absence of any other available information.

\item When comparing our results with the EW($\lambda$5797) - $E(B-V)$ relation for other galaxies, the correlation between these two quantities is revealed. The influence of a second parameter in this relation seems less marked here as for the case of the EW($\lambda$5780) - $E(B-V)$ relation.

\item The distributions of UIBs in the mid-IR and DIBs are similar but with some differences. The largest ones are in the area of the highest CO emission. We preferentially attribute these differences to extinction effects rather than necessarily implying a very different nature of the UIB and DIB carriers. 

\end{enumerate}

The present work demonstrates how IFSs are an efficient tool to collect in one shot and in a relatively short time
plenty of information about DIBs in galaxies other than our own.
With enough spectral resolution, the working philosophy could also be applied to other ISM absorption features.
The itemised conclusions of this work illustrate how the construction of 2D maps for DIBs can be used to better understand the relationship between these features and other components of the ISM in galaxies.
In Section \ref{sec:disc}, we showed that DIB distributions present good -- but not perfect -- correlations with other constituents of the ISM and some possible explanations for these differences were discussed.
An open issue in the comparisons presented in this work is the consequence of the -- surely existing -- projection effects.
Given the violent nature of the interaction in the system, there is a strong possibility that the stars and gas are not well mixed, which in turns could also partially explain the different distributions mapped for the DIBs, $E(B-V)$ and \ion{H}{i}, and UIBs.
From the kinematic point of view, the individual galaxies still display  relatively regular velocity fields (see Paper I, for the ionised gas, Weilbacher et al. in prep, for the stars), with the exception precisely in the overlap region where attenuation high.
Since i) this the first paper that uses IFS-based data to map DIBs in a situation of non-resolved stellar population, ii) a merging system is not the easiest case to start with a thorough discussion about projection effects, we have focused here on the methodology to extract the information and on the possibilities that it can bring, in a more generic manner, to emphasise the potential of the technique.
Specifically, we focused on two of the strongest DIBs and their relation with extinction, atomic and molecular gas, and the PAHs as traced by the UIBs in the mid-infrared.
Similarly, one could explore the association (or lack of) with other components and characteristics of the ISM.
For example, one may wonder how these (and other) DIBs depend on the dust-to-gas ratio, the radiation field, or the gas metallicity  and relative chemical abundances.
The uncertainty of the current data prevent us from exploring these relations. However higher quality data (i.e. deeper and at higher spatial resolution as those that will be provided by, for example, HARMONI at the E-ELT will be instrumental for these purposes.
Learning about these relationships will certainly be key to evaluate the diagnostic potential of these features and will help to put constrains on the nature of the carriers.

\begin{acknowledgements}
We thank Christine Wilson for making
the Caltech Millimeter Array CO map available to us.
We also thank the referee
for valuable and supportive comments that have significantly helped us to improve the first submitted
version of this paper.
AMI acknowledges support from the Spanish MINECO through project
AYA2015-68217-P.
PMW received support through BMBF Verbundforschung (projects MUSE-AO, grant 05A14BAC, and MUSE-NFM, grant 05A17BAA).
Based on observations collected at the European Organisation for Astronomical Research in the Southern Hemisphere under ESO programmes 095.B-0042, 096.B-0017, and 097.B-0346.
This research has made use of the NASA/IPAC Extragalactic Database (NED) which is operated by the Jet Propulsion Laboratory, California Institute of Technology, under contract with the National Aeronautics and Space Administration.
This paper uses the plotting package jmaplot developed by Jes\'us Ma\'{\i}z-Apell\'aniz, \footnote{\texttt{http://jmaiz.iaa.es/software/jmaplot/current/html/ jmaplot\_overview.html}}.
We are also grateful to the communities who developed the many python packages used in this research, such MPDAF \citep{Piqueras17}, Astropy \citep{AstropyCollaboration13}, numpy \citep{Walt11}, scipy \citep{Jones01} and matplotlib \citep{Hunter07}.

\end{acknowledgements}


\bibliographystyle{aa}
\bibliography{mybib_aa.bib}

\begin{thebibliography}{113}
\expandafter\ifx\csname natexlab\endcsname\relax\def\natexlab#1{#1}\fi

\bibitem[{{Asplund} {et~al.}(2009){Asplund}, {Grevesse}, {Sauval}, \&
  {Scott}}]{Asplund09}
{Asplund}, M., {Grevesse}, N., {Sauval}, A.~J., \& {Scott}, P. 2009, \araa, 47,
  481

\bibitem[{{Astropy Collaboration} {et~al.}(2013){Astropy Collaboration},
  {Robitaille}, {Tollerud}, {Greenfield}, {Droettboom}, {Bray}, {Aldcroft},
  {Davis}, {Ginsburg}, {Price-Whelan}, {Kerzendorf}, {Conley}, {Crighton},
  {Barbary}, {Muna}, {Ferguson}, {Grollier}, {Parikh}, {Nair}, {Unther},
  {Deil}, {Woillez}, {Conseil}, {Kramer}, {Turner}, {Singer}, {Fox}, {Weaver},
  {Zabalza}, {Edwards}, {Azalee Bostroem}, {Burke}, {Casey}, {Crawford},
  {Dencheva}, {Ely}, {Jenness}, {Labrie}, {Lim}, {Pierfederici}, {Pontzen},
  {Ptak}, {Refsdal}, {Servillat}, \& {Streicher}}]{AstropyCollaboration13}
{Astropy Collaboration}, {Robitaille}, T.~P., {Tollerud}, E.~J., {et~al.} 2013,
  \aap, 558, A33

\bibitem[{{Bacon} {et~al.}(2012){Bacon}, {Accardo}, {Adjali}, {Anwand},
  {Bauer}, {Blaizot}, {Boudon}, {Brinchmann}, {Brotons}, {Caillier}, {Capoani},
  {Carollo}, {Comin}, {Contini}, {Cumani}, {Daguis}, {Deiries}, {Delabre},
  {Dreizler}, {Dubois}, {Dupieux}, {Dupuy}, {Emsellem}, {Fleischmann}, {Fran{\c
  c}ois}, {Gallou}, {Gharsa}, {Girard}, {Glindemann}, {Guiderdoni}, {Hahn},
  {Hansali}, {Hofmann}, {Jarno}, {Kelz}, {Kiekebusch}, {Knudstrup}, {Koehler},
  {Kollatschny}, {Kosmalski}, {Laurent}, {Le Floch}, {Lilly}, {Lizon {\`a}
  L'Allemand}, {Loupias}, {Manescau}, {Monstein}, {Nicklas}, {Niemeyer},
  {Olaya}, {Palsa}, {Par{\`e}s}, {Pasquini}, {P{\'e}contal-Rousset}, {Pello},
  {Petit}, {Piqueras}, {Popow}, {Reiss}, {Remillieux}, {Renault}, {Rhode},
  {Richard}, {Roth}, {Rupprecht}, {Schaye}, {Slezak}, {Soucail}, {Steinmetz},
  {Streicher}, {Stuik}, {Valentin}, {Vernet}, {Weilbacher}, {Wisotzki},
  {Yerle}, \& {Zins}}]{Bacon12}
{Bacon}, R., {Accardo}, M., {Adjali}, L., {et~al.} 2012, The Messenger, 147, 4

\bibitem[{{Bailey} {et~al.}(2015){Bailey}, {van Loon}, {Sarre}, \&
  {Beckman}}]{Bailey15}
{Bailey}, M., {van Loon}, J.~T., {Sarre}, P.~J., \& {Beckman}, J.~E. 2015,
  \mnras, 454, 4013

\bibitem[{{Baron} {et~al.}(2015){Baron}, {Poznanski}, {Watson}, {Yao}, {Cox},
  \& {Prochaska}}]{Baron15b}
{Baron}, D., {Poznanski}, D., {Watson}, D., {et~al.} 2015, \mnras, 451, 332

\bibitem[{{Bhatt} \& {Cami}(2015)}]{Bhatt15}
{Bhatt}, N.~H. \& {Cami}, J. 2015, \apjs, 216, 22

\bibitem[{{Blades} \& {Madore}(1979)}]{Blades79}
{Blades}, J.~C. \& {Madore}, B.~F. 1979, \aap, 71, 359

\bibitem[{{Brandl} {et~al.}(2005){Brandl}, {Clark}, {Eikenberry}, {Wilson},
  {Henderson}, {Barry}, {Houck}, {Carson}, \& {Hayward}}]{Brandl05}
{Brandl}, B.~R., {Clark}, D.~M., {Eikenberry}, S.~S., {et~al.} 2005, \apj, 635,
  280

\bibitem[{{Brandl} {et~al.}(2009){Brandl}, {Snijders}, {den Brok}, {Whelan},
  {Groves}, {van der Werf}, {Charmandaris}, {Smith}, {Armus}, {Kennicutt}, \&
  {Houck}}]{Brandl09}
{Brandl}, B.~R., {Snijders}, L., {den Brok}, M., {et~al.} 2009, \apj, 699, 1982

\bibitem[{{Bruzual} \& {Charlot}(2003)}]{bru03}
{Bruzual}, G. \& {Charlot}, S. 2003, \mnras, 344, 1000

\bibitem[{{Calzetti} {et~al.}(1997){Calzetti}, {Meurer}, {Bohlin}, {Garnett},
  {Kinney}, {Leitherer}, \& {Storchi-Bergmann}}]{Calzetti97}
{Calzetti}, D., {Meurer}, G.~R., {Bohlin}, R.~C., {et~al.} 1997, \aj, 114, 1834

\bibitem[{{Cami} {et~al.}(1997){Cami}, {Sonnentrucker}, {Ehrenfreund}, \&
  {Foing}}]{Cami97}
{Cami}, J., {Sonnentrucker}, P., {Ehrenfreund}, P., \& {Foing}, B.~H. 1997,
  \aap, 326, 822

\bibitem[{{Campbell} {et~al.}(2015){Campbell}, {Holz}, {Gerlich}, \&
  {Maier}}]{Campbell15}
{Campbell}, E.~K., {Holz}, M., {Gerlich}, D., \& {Maier}, J.~P. 2015, \nat,
  523, 322

\bibitem[{{Capitanio} {et~al.}(2017){Capitanio}, {Lallement}, {Vergely},
  {Elyajouri}, \& {Monreal-Ibero}}]{Capitanio17}
{Capitanio}, L., {Lallement}, R., {Vergely}, J.~L., {Elyajouri}, M., \&
  {Monreal-Ibero}, A. 2017, \aap, 606, A65

\bibitem[{{Cappellari} \& {Copin}(2003)}]{Cappellari03}
{Cappellari}, M. \& {Copin}, Y. 2003, \mnras, 342, 345

\bibitem[{{Cardelli} {et~al.}(1989){Cardelli}, {Clayton}, \& {Mathis}}]{car89}
{Cardelli}, J.~A., {Clayton}, G.~C., \& {Mathis}, J.~S. 1989, \apj, 345, 245

\bibitem[{{Cid Fernandes} {et~al.}(2005){Cid Fernandes}, {Mateus}, {Sodr{\'e}},
  {Stasi{\'n}ska}, \& {Gomes}}]{cid05}
{Cid Fernandes}, R., {Mateus}, A., {Sodr{\'e}}, L., {Stasi{\'n}ska}, G., \&
  {Gomes}, J.~M. 2005, \mnras, 358, 363

\bibitem[{{Cid Fernandes} {et~al.}(2009){Cid Fernandes}, {Schoenell}, {Gomes},
  {Asari}, {Schlickmann}, {Mateus}, {Stasinska}, {Sodr{\'e}}, \&
  {Torres-Papaqui}}]{cid09}
{Cid Fernandes}, R., {Schoenell}, W., {Gomes}, J.~M., {et~al.} 2009, in Revista
  Mexicana de Astronomia y Astrofisica Conference Series, Vol.~35, 127--132

\bibitem[{{Cordiner}(2014)}]{Cordiner14}
{Cordiner}, M.~A. 2014, in IAU Symposium, Vol. 297, The Diffuse Interstellar
  Bands, ed. J.~{Cami} \& N.~L.~J. {Cox}, 41--50

\bibitem[{{Cordiner} {et~al.}(2011){Cordiner}, {Cox}, {Evans}, {Trundle},
  {Smith}, {Sarre}, \& {Gordon}}]{Cordiner11}
{Cordiner}, M.~A., {Cox}, N.~L.~J., {Evans}, C.~J., {et~al.} 2011, \apj, 726,
  39

\bibitem[{{Cordiner} {et~al.}(2008){Cordiner}, {Cox}, {Trundle}, {Evans},
  {Hunter}, {Przybilla}, {Bresolin}, \& {Salama}}]{Cordiner08a}
{Cordiner}, M.~A., {Cox}, N.~L.~J., {Trundle}, C., {et~al.} 2008, \aap, 480,
  L13

\bibitem[{{Cordiner} {et~al.}(2013){Cordiner}, {Fossey}, {Smith}, \&
  {Sarre}}]{Cordiner13}
{Cordiner}, M.~A., {Fossey}, S.~J., {Smith}, A.~M., \& {Sarre}, P.~J. 2013,
  \apjl, 764, L10

\bibitem[{{Cox} {et~al.}(2017){Cox}, {Cami}, {Farhang}, {Smoker},
  {Monreal-Ibero}, {Lallement}, {Sarre}, {Marshall}, {Smith}, {Evans}, {Royer},
  {Linnartz}, {Cordiner}, {Joblin}, {van Loon}, {Foing}, {Bhatt}, {Bron},
  {Elyajouri}, {de Koter}, {Ehrenfreund}, {Javadi}, {Kaper}, {Khosroshadi},
  {Laverick}, {Le Petit}, {Mulas}, {Roueff}, {Salama}, \& {Spaans}}]{Cox17}
{Cox}, N., {Cami}, J., {Farhang}, A., {et~al.} 2017, ArXiv e-prints
  [\eprint[arXiv]{1708.01429}]

\bibitem[{{Cox} {et~al.}(2014){Cox}, {Cami}, {Kaper}, {Ehrenfreund}, {Foing},
  {Ochsendorf}, {van Hooff}, \& {Salama}}]{Cox14}
{Cox}, N.~L.~J., {Cami}, J., {Kaper}, L., {et~al.} 2014, \aap, 569, A117

\bibitem[{{Cox} {et~al.}(2006){Cox}, {Cordiner}, {Cami}, {Foing}, {Sarre},
  {Kaper}, \& {Ehrenfreund}}]{Cox06}
{Cox}, N.~L.~J., {Cordiner}, M.~A., {Cami}, J., {et~al.} 2006, \aap, 447, 991

\bibitem[{{Cox} {et~al.}(2007){Cox}, {Cordiner}, {Ehrenfreund}, {Kaper},
  {Sarre}, {Foing}, {Spaans}, {Cami}, {Sofia}, {Clayton}, {Gordon}, \&
  {Salama}}]{Cox07}
{Cox}, N.~L.~J., {Cordiner}, M.~A., {Ehrenfreund}, P., {et~al.} 2007, \aap,
  470, 941

\bibitem[{{Cox} \& {Patat}(2008)}]{Cox08}
{Cox}, N.~L.~J. \& {Patat}, F. 2008, \aap, 485, L9

\bibitem[{{Cox} \& {Spaans}(2006)}]{Cox06b}
{Cox}, N.~L.~J. \& {Spaans}, M. 2006, \aap, 451, 973

\bibitem[{{D'Odorico} {et~al.}(1989){D'Odorico}, {di Serego Alighieri},
  {Pettini}, {Magain}, {Nissen}, \& {Panagia}}]{DOdorico89}
{D'Odorico}, S., {di Serego Alighieri}, S., {Pettini}, M., {et~al.} 1989, \aap,
  215, 21

\bibitem[{{Draine} \& {Li}(2007)}]{Draine07}
{Draine}, B.~T. \& {Li}, A. 2007, \apj, 657, 810

\bibitem[{{Ehrenfreund} {et~al.}(2002){Ehrenfreund}, {Cami},
  {Jim{\'e}nez-Vicente}, {Foing}, {Kaper}, {van der Meer}, {Cox},
  {D'Hendecourt}, {Maier}, {Salama}, {Sarre}, {Snow}, \&
  {Sonnentrucker}}]{Ehrenfreund02}
{Ehrenfreund}, P., {Cami}, J., {Jim{\'e}nez-Vicente}, J., {et~al.} 2002, \apjl,
  576, L117

\bibitem[{{Elyajouri} {et~al.}(2017){Elyajouri}, {Lallement}, {Monreal-Ibero},
  {Capitanio}, \& {Cox}}]{Elyajouri17}
{Elyajouri}, M., {Lallement}, R., {Monreal-Ibero}, A., {Capitanio}, L., \&
  {Cox}, N.~L.~J. 2017, \aap, 600, A129

\bibitem[{{Ensor} {et~al.}(2017){Ensor}, {Cami}, {Bhatt}, \& {Soddu}}]{Ensor17}
{Ensor}, T., {Cami}, J., {Bhatt}, N.~H., \& {Soddu}, A. 2017, \apj, 836, 162

\bibitem[{{Fluks} {et~al.}(1994){Fluks}, {Plez}, {The}, {de Winter},
  {Westerlund}, \& {Steenman}}]{Fluks94}
{Fluks}, M.~A., {Plez}, B., {The}, P.~S., {et~al.} 1994, \aaps, 105, 311

\bibitem[{{Friedman} {et~al.}(2011){Friedman}, {York}, {McCall}, {Dahlstrom},
  {Sonnentrucker}, {Welty}, {Drosback}, {Hobbs}, {Rachford}, \&
  {Snow}}]{Friedman11}
{Friedman}, S.~D., {York}, D.~G., {McCall}, B.~J., {et~al.} 2011, \apj, 727, 33

\bibitem[{{Galazutdinov} \& {Kre{\l}owski}(2017)}]{Galazutdinov17}
{Galazutdinov}, G.~A. \& {Kre{\l}owski}, J. 2017, \actaa, 67, 159

\bibitem[{{Geballe} {et~al.}(2011){Geballe}, {Najarro}, {Figer},
  {Schlegelmilch}, \& {de La Fuente}}]{Geballe11}
{Geballe}, T.~R., {Najarro}, F., {Figer}, D.~F., {Schlegelmilch}, B.~W., \& {de
  La Fuente}, D. 2011, \nat, 479, 200

\bibitem[{{Girardi} {et~al.}(2000){Girardi}, {Bressan}, {Bertelli}, \&
  {Chiosi}}]{gir00}
{Girardi}, L., {Bressan}, A., {Bertelli}, G., \& {Chiosi}, C. 2000, \aaps, 141,
  371

\bibitem[{{Gordon} {et~al.}(2003){Gordon}, {Clayton}, {Misselt}, {Landolt}, \&
  {Wolff}}]{Gordon03}
{Gordon}, K.~D., {Clayton}, G.~C., {Misselt}, K.~A., {Landolt}, A.~U., \&
  {Wolff}, M.~J. 2003, \apj, 594, 279

\bibitem[{{Hamano} {et~al.}(2016){Hamano}, {Kobayashi}, {Kondo}, {Sameshima},
  {Nakanishi}, {Ikeda}, {Yasui}, {Mizumoto}, {Matsunaga}, {Fukue}, {Yamamoto},
  {Izumi}, {Mito}, {Nakaoka}, {Kawanishi}, {Kitano}, {Otsubo}, {Kinoshita}, \&
  {Kawakita}}]{Hamano16}
{Hamano}, S., {Kobayashi}, N., {Kondo}, S., {et~al.} 2016, \apj, 821, 42

\bibitem[{{Heckman} \& {Lehnert}(2000)}]{Heckman00}
{Heckman}, T.~M. \& {Lehnert}, M.~D. 2000, \apj, 537, 690

\bibitem[{{Heger}(1922)}]{Heger22}
{Heger}, M.~L. 1922, Lick Observatory Bulletin, 10, 141

\bibitem[{{Herbig}(1993)}]{Herbig93}
{Herbig}, G.~H. 1993, \apj, 407, 142

\bibitem[{{Herbig}(1995)}]{Herbig95}
{Herbig}, G.~H. 1995, \araa, 33, 19

\bibitem[{{Hibbard} {et~al.}(2001){Hibbard}, {van der Hulst}, {Barnes}, \&
  {Rich}}]{Hibbard01}
{Hibbard}, J.~E., {van der Hulst}, J.~M., {Barnes}, J.~E., \& {Rich}, R.~M.
  2001, \aj, 122, 2969

\bibitem[{{Hobbs} {et~al.}(2009){Hobbs}, {York}, {Thorburn}, {Snow}, {Bishof},
  {Friedman}, {McCall}, {Oka}, {Rachford}, {Sonnentrucker}, \&
  {Welty}}]{Hobbs09}
{Hobbs}, L.~M., {York}, D.~G., {Thorburn}, J.~A., {et~al.} 2009, \apj, 705, 32

\bibitem[{Hunter(2007)}]{Hunter07}
Hunter, J.~D. 2007, Computing In Science \& Engineering, 9, 90

\bibitem[{{Hutchings}(1966)}]{Hutchings66}
{Hutchings}, J.~B. 1966, \mnras, 131, 299

\bibitem[{{Iglesias-Groth}(2007)}]{IglesiasGroth07}
{Iglesias-Groth}, S. 2007, \apjl, 661, L167

\bibitem[{{Jenniskens} \& {Desert}(1994)}]{Jenniskens94}
{Jenniskens}, P. \& {Desert}, F.-X. 1994, \aaps, 106

\bibitem[{{Joblin} {et~al.}(1990){Joblin}, {D'Hendecourt}, {Leger}, \&
  {Maillard}}]{Joblin90}
{Joblin}, C., {D'Hendecourt}, L., {Leger}, A., \& {Maillard}, J.~P. 1990, \nat,
  346, 729

\bibitem[{Jones {et~al.}(2001--)Jones, Oliphant, Peterson, {et~al.}}]{Jones01}
Jones, E., Oliphant, T., Peterson, P., {et~al.} 2001--, {SciPy}: Open source
  scientific tools for {Python}

\bibitem[{{Junkkarinen} {et~al.}(2004){Junkkarinen}, {Cohen}, {Beaver},
  {Burbidge}, {Lyons}, \& {Madejski}}]{Junkkarinen04}
{Junkkarinen}, V.~T., {Cohen}, R.~D., {Beaver}, E.~A., {et~al.} 2004, \apj,
  614, 658

\bibitem[{{Kos}(2017)}]{Kos17}
{Kos}, J. 2017, \mnras, 468, 4255

\bibitem[{{Kos} {et~al.}(2013){Kos}, {Zwitter}, {Grebel}, {Bienayme}, {Binney},
  {Bland-Hawthorn}, {Freeman}, {Gibson}, {Gilmore}, {Kordopatis}, {Navarro},
  {Parker}, {Reid}, {Seabroke}, {Siebert}, {Siviero}, {Steinmetz}, {Watson}, \&
  {Wyse}}]{Kos13}
{Kos}, J., {Zwitter}, T., {Grebel}, E.~K., {et~al.} 2013, \apj, 778, 86

\bibitem[{{Kre{\l}owski} {et~al.}(2011){Kre{\l}owski}, {Galazutdinov}, \&
  {Ko{\l}os}}]{Krewlowski11}
{Kre{\l}owski}, J., {Galazutdinov}, G., \& {Ko{\l}os}, R. 2011, \apj, 735, 124

\bibitem[{{Krelowski} {et~al.}(1992){Krelowski}, {Snow}, {Seab}, \&
  {Papaj}}]{Krelowski92}
{Krelowski}, J., {Snow}, T.~P., {Seab}, C.~G., \& {Papaj}, J. 1992, \mnras,
  258, 693

\bibitem[{{Kroto}(1988)}]{Kroto88}
{Kroto}, H. 1988, Science, 242, 1139

\bibitem[{{Lan} {et~al.}(2015){Lan}, {M{\'e}nard}, \& {Zhu}}]{Lan15}
{Lan}, T.-W., {M{\'e}nard}, B., \& {Zhu}, G. 2015, \mnras, 452, 3629

\bibitem[{{Lardo} {et~al.}(2015){Lardo}, {Davies}, {Kudritzki}, {Gazak},
  {Evans}, {Patrick}, {Bergemann}, \& {Plez}}]{Lardo15}
{Lardo}, C., {Davies}, B., {Kudritzki}, R.-P., {et~al.} 2015, \apj, 812, 160

\bibitem[{{Leger} \& {D'Hendecourt}(1985)}]{Leger85}
{Leger}, A. \& {D'Hendecourt}, L. 1985, \aap, 146, 81

\bibitem[{{Maier} {et~al.}(2004){Maier}, {Walker}, \& {Bohlender}}]{Maier04}
{Maier}, J.~P., {Walker}, G.~A.~H., \& {Bohlender}, D.~A. 2004, \apj, 602, 286

\bibitem[{{Markwardt}(2009)}]{Markwardt09}
{Markwardt}, C.~B. 2009, in Astronomical Society of the Pacific Conference
  Series, ed. {D.~A.~Bohlender, D.~Durand, \& P.~Dowler}, Vol. 411, 251

\bibitem[{{Merrill}(1936)}]{Merrill36}
{Merrill}, P.~W. 1936, \apj, 83, 126

\bibitem[{{Mirabel} {et~al.}(1998){Mirabel}, {Vigroux}, {Charmandaris},
  {Sauvage}, {Gallais}, {Tran}, {Cesarsky}, {Madden}, \& {Duc}}]{Mirabel98}
{Mirabel}, I.~F., {Vigroux}, L., {Charmandaris}, V., {et~al.} 1998, \aap, 333,
  L1

\bibitem[{{Monreal-Ibero} {et~al.}(2015{\natexlab{a}}){Monreal-Ibero},
  {Lallement}, {Puspitarini}, {Bonifacio}, \& {Monaco}}]{MonrealIbero15b}
{Monreal-Ibero}, A., {Lallement}, R., {Puspitarini}, L., {Bonifacio}, P., \&
  {Monaco}, L. 2015{\natexlab{a}}, \memsai, 86, 527

\bibitem[{{Monreal-Ibero} {et~al.}(2010){Monreal-Ibero}, {V{\'{\i}}lchez},
  {Walsh}, \& {Mu{\~n}oz-Tu{\~n}{\'o}n}}]{MonrealIbero10a}
{Monreal-Ibero}, A., {V{\'{\i}}lchez}, J.~M., {Walsh}, J.~R., \&
  {Mu{\~n}oz-Tu{\~n}{\'o}n}, C. 2010, \aap, 517, A27+

\bibitem[{{Monreal-Ibero} {et~al.}(2015{\natexlab{b}}){Monreal-Ibero},
  {Weilbacher}, {Wendt}, {Selman}, {Lallement}, {Brinchmann}, {Kamann}, \&
  {Sandin}}]{MonrealIbero15}
{Monreal-Ibero}, A., {Weilbacher}, P.~M., {Wendt}, M., {et~al.}
  2015{\natexlab{b}}, \aap, 576, L3

\bibitem[{{Motylewski} {et~al.}(2000){Motylewski}, {Linnartz}, {Vaizert},
  {Maier}, {Galazutdinov}, {Musaev}, {Kre{\l}owski}, {Walker}, \&
  {Bohlender}}]{Motylewski00}
{Motylewski}, T., {Linnartz}, H., {Vaizert}, O., {et~al.} 2000, \apj, 531, 312

\bibitem[{{Osterbrock} \& {Ferland}(2006)}]{Osterbrock06}
{Osterbrock}, D.~E. \& {Ferland}, G.~J. 2006, {Astrophysics of gaseous nebulae
  and active galactic nuclei}, ed. D.~E. {Osterbrock} \& G.~J. {Ferland}

\bibitem[{{Phillips} {et~al.}(2013){Phillips}, {Simon}, {Morrell}, {Burns},
  {Cox}, {Foley}, {Karakas}, {Patat}, {Sternberg}, {Williams}, {Gal-Yam},
  {Hsiao}, {Leonard}, {Persson}, {Stritzinger}, {Thompson}, {Campillay},
  {Contreras}, {Folatelli}, {Freedman}, {Hamuy}, {Roth}, {Shields}, {Suntzeff},
  {Chomiuk}, {Ivans}, {Madore}, {Penprase}, {Perley}, {Pignata}, {Preston}, \&
  {Soderberg}}]{Phillips13}
{Phillips}, M.~M., {Simon}, J.~D., {Morrell}, N., {et~al.} 2013, \apj, 779, 38

\bibitem[{{Piqueras} {et~al.}(2017){Piqueras}, {Conseil}, {Shepherd}, {Bacon},
  {Leclercq}, \& {Richard}}]{Piqueras17}
{Piqueras}, L., {Conseil}, S., {Shepherd}, M., {et~al.} 2017, to be published
  in ADASS XXVI, ArXiv e-prints [\eprint[arXiv]{1710.03554}]

\bibitem[{{Puspitarini} {et~al.}(2015){Puspitarini}, {Lallement}, {Babusiaux},
  {Chen}, {Bonifacio}, {Sbordone}, {Caffau}, {Duffau}, {Hill}, {Monreal-Ibero},
  {Royer}, {Arenou}, {Peralta}, {Drew}, {Bonito}, {Lopez-Santiago}, {Alfaro},
  {Bensby}, {Bragaglia}, {Flaccomio}, {Lanzafame}, {Pancino}, {Recio-Blanco},
  {Smiljanic}, {Costado}, {Lardo}, {de Laverny}, \& {Zwitter}}]{Puspitarini15}
{Puspitarini}, L., {Lallement}, R., {Babusiaux}, C., {et~al.} 2015, \aap, 573,
  A35

\bibitem[{{Puspitarini} {et~al.}(2013){Puspitarini}, {Lallement}, \&
  {Chen}}]{Puspitarini13}
{Puspitarini}, L., {Lallement}, R., \& {Chen}, H.-C. 2013, \aap, 555, A25

\bibitem[{{Rieke} \& {Lebofsky}(1985)}]{Rieke85}
{Rieke}, G.~H. \& {Lebofsky}, M.~J. 1985, \apj, 288, 618

\bibitem[{{Rigopoulou} {et~al.}(1999){Rigopoulou}, {Spoon}, {Genzel}, {Lutz},
  {Moorwood}, \& {Tran}}]{Rigopoulou99}
{Rigopoulou}, D., {Spoon}, H.~W.~W., {Genzel}, R., {et~al.} 1999, \aj, 118,
  2625

\bibitem[{{Ritchey} \& {Wallerstein}(2015)}]{Ritchey15}
{Ritchey}, A.~M. \& {Wallerstein}, G. 2015, \pasp, 127, 223

\bibitem[{{Russell} \& {Dopita}(1992)}]{Russell92}
{Russell}, S.~C. \& {Dopita}, M.~A. 1992, \apj, 384, 508

\bibitem[{{Salama} {et~al.}(1996){Salama}, {Bakes}, {Allamandola}, \&
  {Tielens}}]{Salama96}
{Salama}, F., {Bakes}, E.~L.~O., {Allamandola}, L.~J., \& {Tielens},
  A.~G.~G.~M. 1996, \apj, 458, 621

\bibitem[{{Sanders} {et~al.}(2003){Sanders}, {Mazzarella}, {Kim}, {Surace}, \&
  {Soifer}}]{Sanders03}
{Sanders}, D.~B., {Mazzarella}, J.~M., {Kim}, D.-C., {Surace}, J.~A., \&
  {Soifer}, B.~T. 2003, \aj, 126, 1607

\bibitem[{{Sanders} {et~al.}(2012){Sanders}, {Caldwell}, {McDowell}, \&
  {Harding}}]{Sanders12}
{Sanders}, N.~E., {Caldwell}, N., {McDowell}, J., \& {Harding}, P. 2012, \apj,
  758, 133

\bibitem[{{Sarre}(2006)}]{Sarre06}
{Sarre}, P.~J. 2006, Journal of Molecular Spectroscopy, 238, 1

\bibitem[{{Sassara} {et~al.}(2001){Sassara}, {Zerza}, {Chergui}, \&
  {Leach}}]{Sassara01}
{Sassara}, A., {Zerza}, G., {Chergui}, M., \& {Leach}, S. 2001, \apjs, 135, 263

\bibitem[{{Schlafly} \& {Finkbeiner}(2011)}]{Schlafly11}
{Schlafly}, E.~F. \& {Finkbeiner}, D.~P. 2011, \apj, 737, 103

\bibitem[{{Schweizer} {et~al.}(2008){Schweizer}, {Burns}, {Madore}, {Mager},
  {Phillips}, {Freedman}, {Boldt}, {Contreras}, {Folatelli}, {Gonz{\'a}lez},
  {Hamuy}, {Krzeminski}, {Morrell}, {Persson}, {Roth}, \&
  {Stritzinger}}]{Schweizer08}
{Schweizer}, F., {Burns}, C.~R., {Madore}, B.~F., {et~al.} 2008, \aj, 136, 1482

\bibitem[{{Smith} {et~al.}(2007){Smith}, {Struck}, {Hancock}, {Appleton},
  {Charmandaris}, \& {Reach}}]{Smith07}
{Smith}, B.~J., {Struck}, C., {Hancock}, M., {et~al.} 2007, \aj, 133, 791

\bibitem[{{Snow}(2014)}]{Snow14}
{Snow}, T.~P. 2014, in IAU Symposium, Vol. 297, The Diffuse Interstellar Bands,
  ed. J.~{Cami} \& N.~L.~J. {Cox}, 3--12

\bibitem[{{Snow} \& {Cohen}(1974)}]{Snow74}
{Snow}, Jr., T.~P. \& {Cohen}, J.~G. 1974, \apj, 194, 313

\bibitem[{{Sollerman} {et~al.}(2005){Sollerman}, {Cox}, {Mattila},
  {Ehrenfreund}, {Kaper}, {Leibundgut}, \& {Lundqvist}}]{Sollerman05}
{Sollerman}, J., {Cox}, N., {Mattila}, S., {et~al.} 2005, \aap, 429, 559

\bibitem[{{Srianand} {et~al.}(2013){Srianand}, {Gupta}, {Rahmani}, {Momjian},
  {Petitjean}, \& {Noterdaeme}}]{Srianand13}
{Srianand}, R., {Gupta}, N., {Rahmani}, H., {et~al.} 2013, \mnras, 428, 2198

\bibitem[{{Th{\"o}ne} {et~al.}(2009){Th{\"o}ne}, {Micha{\l}owski}, {Leloudas},
  {Cox}, {Fynbo}, {Sollerman}, {Hjorth}, \& {Vreeswijk}}]{Thoene09}
{Th{\"o}ne}, C.~C., {Micha{\l}owski}, M.~J., {Leloudas}, G., {et~al.} 2009,
  \apj, 698, 1307

\bibitem[{{Toribio San Cipriano} {et~al.}(2016){Toribio San Cipriano},
  {Garc{\'{\i}}a-Rojas}, {Esteban}, {Bresolin}, \&
  {Peimbert}}]{ToribioSanCipriano16}
{Toribio San Cipriano}, L., {Garc{\'{\i}}a-Rojas}, J., {Esteban}, C.,
  {Bresolin}, F., \& {Peimbert}, M. 2016, \mnras, 458, 1866

\bibitem[{{van der Zwet} \& {Allamandola}(1985)}]{vanderZwet85}
{van der Zwet}, G.~P. \& {Allamandola}, L.~J. 1985, \aap, 146, 76

\bibitem[{{van Loon} {et~al.}(2013){van Loon}, {Bailey}, {Tatton}, {Ma{\'{\i}}z
  Apell{\'a}niz}, {Crowther}, {de Koter}, {Evans}, {H{\'e}nault-Brunet},
  {Howarth}, {Richter}, {Sana}, {Sim{\'o}n-D{\'{\i}}az}, {Taylor}, \&
  {Walborn}}]{vanLoon13}
{van Loon}, J.~T., {Bailey}, M., {Tatton}, B.~L., {et~al.} 2013, \aap, 550,
  A108

\bibitem[{{van Loon} {et~al.}(2009){van Loon}, {Smith}, {McDonald}, {Sarre},
  {Fossey}, \& {Sharp}}]{vanLoon09}
{van Loon}, J.~T., {Smith}, K.~T., {McDonald}, I., {et~al.} 2009, \mnras, 399,
  195

\bibitem[{{Vigroux} {et~al.}(1996){Vigroux}, {Mirabel}, {Altieri}, {Boulanger},
  {Cesarsky}, {Cesarsky}, {Claret}, {Fransson}, {Gallais}, {Levine}, {Madden},
  {Okumura}, \& {Tran}}]{Vigroux96}
{Vigroux}, L., {Mirabel}, F., {Altieri}, B., {et~al.} 1996, \aap, 315, L93

\bibitem[{{Vos} {et~al.}(2011){Vos}, {Cox}, {Kaper}, {Spaans}, \&
  {Ehrenfreund}}]{Vos11}
{Vos}, D.~A.~I., {Cox}, N.~L.~J., {Kaper}, L., {Spaans}, M., \& {Ehrenfreund},
  P. 2011, \aap, 533, A129

\bibitem[{{Walker} {et~al.}(2015){Walker}, {Bohlender}, {Maier}, \&
  {Campbell}}]{Walker15}
{Walker}, G.~A.~H., {Bohlender}, D.~A., {Maier}, J.~P., \& {Campbell}, E.~K.
  2015, \apjl, 812, L8

\bibitem[{Walt {et~al.}(2011)Walt, Colbert, \& Varoquaux}]{Walt11}
Walt, S. v.~d., Colbert, S.~C., \& Varoquaux, G. 2011, Computing in Science and
  Engg., 13, 22

\bibitem[{{Wang} {et~al.}(2004){Wang}, {Fazio}, {Ashby}, {Huang}, {Pahre},
  {Smith}, {Willner}, {Forrest}, {Pipher}, \& {Surace}}]{Wang04}
{Wang}, Z., {Fazio}, G.~G., {Ashby}, M.~L.~N., {et~al.} 2004, \apjs, 154, 193

\bibitem[{{Weilbacher} {et~al.}(2017){Weilbacher}, {Monreal-Ibero}, {Verhamme},
  {Sandin}, {Steinmetz}, {Kollatschny}, {Krajnovi{\'c}}, {Kamann}, {Roth},
  {Erroz-Ferrer}, {Marino}, {Maseda}, {Wendt}, {Bacon}, {Dreizler}, {Richard},
  \& {Wisotzki}}]{WMV+17}
{Weilbacher}, P.~M., {Monreal-Ibero}, A., {Verhamme}, A., {et~al.} 2017, ArXiv
  e-prints [\eprint[arXiv]{1712.04450}], ({P}aper~I)

\bibitem[{{Weilbacher} {et~al.}(2014){Weilbacher}, {Streicher}, {Urrutia},
  {P{\'e}contal-Rousset}, {Jarno}, \& {Bacon}}]{2014ASPC..485..451W}
{Weilbacher}, P.~M., {Streicher}, O., {Urrutia}, T., {et~al.} 2014, in
  ASP~Conf.~Ser., Vol. 485, Astronomical Data Analysis Software and Systems
  XXIII, ed. N.~{Manset} \& P.~{Forshay}, 451

\bibitem[{{Welty}(2014)}]{Welty14b}
{Welty}, D.~E. 2014, in IAU Symposium, Vol. 297, The Diffuse Interstellar
  Bands, ed. J.~{Cami} \& N.~L.~J. {Cox}, 153--162

\bibitem[{{Welty} {et~al.}(2006){Welty}, {Federman}, {Gredel}, {Thorburn}, \&
  {Lambert}}]{Welty06}
{Welty}, D.~E., {Federman}, S.~R., {Gredel}, R., {Thorburn}, J.~A., \&
  {Lambert}, D.~L. 2006, \apjs, 165, 138

\bibitem[{{Welty} {et~al.}(2014){Welty}, {Ritchey}, {Dahlstrom}, \&
  {York}}]{Welty14}
{Welty}, D.~E., {Ritchey}, A.~M., {Dahlstrom}, J.~A., \& {York}, D.~G. 2014,
  \apj, 792, 106

\bibitem[{{Wendt} {et~al.}(2017){Wendt}, {Husser}, {Kamann}, {Monreal-Ibero},
  {Richter}, {Brinchmann}, {Dreizler}, {Weilbacher}, \& {Wisotzki}}]{Wendt17}
{Wendt}, M., {Husser}, T.-O., {Kamann}, S., {et~al.} 2017, \aap, 607, A133

\bibitem[{{Weselak} {et~al.}(2004){Weselak}, {Galazutdinov}, {Musaev}, \&
  {Kre{\l}owski}}]{Weselak04}
{Weselak}, T., {Galazutdinov}, G.~A., {Musaev}, F.~A., \& {Kre{\l}owski}, J.
  2004, \aap, 414, 949

\bibitem[{{Whitmore} {et~al.}(2014){Whitmore}, {Brogan}, {Chandar}, {Evans},
  {Hibbard}, {Johnson}, {Leroy}, {Privon}, {Remijan}, \& {Sheth}}]{Whitmore14}
{Whitmore}, B.~C., {Brogan}, C., {Chandar}, R., {et~al.} 2014, \apj, 795, 156

\bibitem[{{Whitmore} {et~al.}(2010){Whitmore}, {Chandar}, {Schweizer},
  {Rothberg}, {Leitherer}, {Rieke}, {Rieke}, {Blair}, {Mengel}, \&
  {Alonso-Herrero}}]{Whitmore10}
{Whitmore}, B.~C., {Chandar}, R., {Schweizer}, F., {et~al.} 2010, \aj, 140, 75

\bibitem[{{Wilson} {et~al.}(2000){Wilson}, {Scoville}, {Madden}, \&
  {Charmandaris}}]{Wilson00}
{Wilson}, C.~D., {Scoville}, N., {Madden}, S.~C., \& {Charmandaris}, V. 2000,
  \apj, 542, 120

\bibitem[{{Worthey} {et~al.}(1994){Worthey}, {Faber}, {Gonzalez}, \&
  {Burstein}}]{Worthey94}
{Worthey}, G., {Faber}, S.~M., {Gonzalez}, J.~J., \& {Burstein}, D. 1994,
  \apjs, 94, 687

\bibitem[{{Zaragoza-Cardiel} {et~al.}(2014){Zaragoza-Cardiel}, {Font},
  {Beckman}, {Garc{\'{\i}}a-Lorenzo}, {Erroz-Ferrer}, \&
  {Guti{\'e}rrez}}]{ZaragozaCardiel14}
{Zaragoza-Cardiel}, J., {Font}, J., {Beckman}, J.~E., {et~al.} 2014, \mnras,
  445, 1412

\bibitem[{{Zhu} {et~al.}(2003){Zhu}, {Seaquist}, \& {Kuno}}]{Zhu03}
{Zhu}, M., {Seaquist}, E.~R., \& {Kuno}, N. 2003, \apj, 588, 243

\end{thebibliography}


\end{document}